\documentclass{mn2e}
\usepackage{times}
\usepackage{graphicx}
\usepackage{soul}       
\usepackage{amsmath}
\usepackage{txfonts}

\def\Msolar{\hbox{${\rm M}_\odot$}}

\begin{document} 

\title[Extinction law in 30\,Dor]{The extinction law inside the
30\,Doradus nebula\thanks{Based on observations with the  NASA/ESA {\it
Hubble Space Telescope}, obtained at the Space Telescope Science
Institute, which is operated by AURA, Inc., under NASA contract
NAS5-26555.}}

\author[Guido De Marchi \& Nino Panagia]
{Guido~De~Marchi$^1$ and Nino~Panagia,$^{2,3,4}$\\
$^1$European Space Research and Technology Centre, Keplerlaan 1, 2200 AG
Noordwijk, The Netherlands, gdemarchi@esa.int \\
$^2$Space Telescope Science Institute, 3700 San Martin Drive, Baltimore, MD
21218, USA, panagia@stsci.edu\\
$^3$INAF--NA, Osservatorio Astronomico di Capodimonte, Salita Moiariello
16, 80131 Naples, Italy\\
$^4$Supernova Ltd, OYV \#131, Northsound Rd., Virgin Gorda VG1150,
Virgin Islands, UK}

\date{Received 13 May 2014; Accepted 18 August 2014}
\pagerange{\pageref{firstpage}--\pageref{lastpage}} \pubyear{2014}

\maketitle

\begin{abstract}

We have studied the interstellar extinction in a field of $\sim 3\arcmin
\times 3\arcmin$ at the core of the 30\,Doradus nebula, including the
central R\,136 cluster, in the Large Magellanic Cloud. Observations at
optical and near-infrared wavelengths, obtained  with the WFC\,3 camera
on board the {\em Hubble Space Telescope}, show that the stars belonging
to the red giant clump are spread across  the colour--magnitude diagrams
because of the considerable and uneven levels of extinction in this
region. Since these stars share very similar physical properties and are
all at the same distance, they allow us to derive the absolute
extinction in a straightforward and reliable way. Thus we have measured
the extinction towards about 180 objects and the extinction law in the
range $0.3-1.6\,\muup$m. At optical wavelengths, the extinction curve is
almost parallel to that of the diffuse Galactic interstellar medium.
Taking the latter as a template, the value of $R_V = 4.5 \pm 0.2$ 
that we measure indicates that in the optical there is an extra grey
component due to a  larger fraction of large grains. At wavelengths
longer than $\sim 1$\,$\muup$m, the contribution of this additional
component tapers off as $\lambda^{-1.5}$, like in the Milky Way,
suggesting that the nature of the grains is otherwise similar to those
in our Galaxy, but with a $\sim 2.2$ times higher fraction of large 
grains. These results are consistent with the addition of ``fresh'' 
large grains by supernova explosions, as recently revealed by
{\em Herschel} and {\em ALMA} observations of SN\,1987A.

\end{abstract}


\section{Introduction}

Studies of the properties of stellar populations hinge on the accurate
knowledge of the amount and properties of interstellar extinction, 
since they affect fundamental observational quantities such as the
distance and luminosity function (e.g. Gottlieb \& Upson 1969). The
diffuse Galactic interstellar medium (ISM) appears to have a rather 
uniform extinction law and the ratio of total and selective extinction 
$R_V = A_V/E(B - V)$ is found to be consistently about $3.1$ (e.g.
Savage \& Mathis 1979). However, in star forming regions the situation
is rather different: $R_V$ varies considerably in our Galaxy  and
studies conducted over the past 40 years have revealed a wide variety of
extinction curves (e.g. Bless \& Savage 1972; Seaton 1979; Fitzpatrick
1998, 1999). In particular, in dense star forming regions the situation
is quite different from that of the diffuse ISM and generally the value
of $R_V$ appears to increase {(e.g. Johnson \& Mendoza 1964; Savage \& 
Mathis 1979; Cardelli, Clayton \& Mathis 1988)}, due to the presence of 
larger grains, with a correspondingly rather different shape of the 
extinction curve. 

{Nonetheless, there appears to be some commonality in the properties of
the extinction in these environments. Even though the extinction curves
are effectively different, Cardelli, Clayton \& Mathis (1989) have shown
that the properties of the extinction at optical and near-infrared (NIR)
wavelengths appear to be correlated with those in the ultraviolet, and
the measured $A(\lambda)/A(V)$ can generally be described by one family
of curves depending only on $R_V$. For instance, Valencic, Clayton \&
Gordon (2004) compiled a homogeneous database of 417 extinction curves
towards Galactic stars, finding that 93\,\% of them are compatible with
the $R_V$-dependent extinction parameterization of Cardelli et al.
(1989), within the uncertainties. Fitzpatrick \& Massa (2007) offer
a different interpretation of the results by Valencic et al. (2004),
pointing out that the dependence of $A(\lambda)/A(V)$ on  $R_V$ only is
partially caused by the fact that $A(\lambda)/A(V)$ is derived from the
measured colour excess by taking the value of $R_V$ into account,
therefore necessarily causing some degree of apparent correlation.
According to Fitzpatrick \& Massa (2007), extinction curves are not a 
one-parameter family in $R_V$ and a variety of curves remains even in
our Galaxy.

In this work we do not address the matter of the functional form of
the extinction laws, but rather concentrate on the physical properties
of the extinction inside the Tarantula nebula (30\,Dor). A robust
determination of the extinction curve in this environment} is crucial
for modern Astronomy, since this object is the closest and only known
starburst in the Local Group. The strength of star formation in 30\,Dor
and the number of massive stars that it harbours are similar to those
found in interacting galaxies in the local Universe and in young
galaxies at high redshift ($z > 5$, e.g. Meurer et al. 1997;  Shapley et
al. 2003; Heckman et al. 2004). Furthermore, with a metallicity of the
order of 1/3 $Z_\odot$ (e.g. Hill, Andrievsky \& Spite 1995; Geha et al.
1998), 30\,Dor allows us to probe the prevailing conditions at redshift
$z \gtrsim 2$, when star formation was at its peak in the Universe (e.g.
Madau et al. 1996; Lilly et al 1996). Thus, determining the properties
of the extinction in 30\,Dor will not  only allow us to understand how
star formation is proceeding in this specific Large Magellanic Cloud
(LMC) environment, but also in a cosmological sense.

{Gordon et al. (2003) have carried out an exhaustive study of the
extinction properties in the LMC towards 19 different lines of sight,
extending previous investigations by Clayton \& Martin (1985) and
Misselt, Clayton \& Gordon al. (1999). They find that only four of the
19 sight lines probed appear to be compatible with the $R_V$-dependent
family of extinction curves of Cardelli et al. (1989), thereby revealing
a wider variety than in the Milky Way. Unfortunately, none of the sight
lines probe specifically the 30\,Dor region: although 8 of them are
associated with the LMC2 Supershell near the Tarantula nebula, the
nearest star is $\sim 9^\prime$ away from the centre of 30\,Dor and the
others are beyond $15^\prime$, with a median distance of $\sim
20^\prime$, typically probing a projected distance of $\sim 300$\,pc or
three times the radius of the nebula (Lebouteiller et al. 2008). }

Until recently, the most accurate empirical determination of the
extinction law inside 30\,Dor was the 30-year old study by Fitzpatrick
\& Savage (1984). These authors used the traditional approach of the
``pair method,'' in which the flux distribution or colours of a reddened
object are compared with those of a reference star of the same spectral
type ({\em e.g.,} Johnson 1968; Massa, Savage \& Fitzpatrick 1983;
Cardelli, Sembach \& Mathis 1992). This method is very reliable and can
detect subtle variations of the extinction law with the environment, but
it works optimally if using targets such as massive stars with high 
quality spectra extending from the near ultraviolet to the NIR. 

Given the level of crowding present inside 30\,Dor, the extinction law
derived by Fitzpatrick \& Savage (1984) is based on just a handful of
lines of sight (namely 7 within the central $5\arcmin$ of R\,136a). This
situation has been recently improved by Ma{\'\i}z Apell{\'a}niz et al.
(2014), who used a Bayesian approach to derive the so-called
``extinction without standards'' (Fitzpatrick \& Massa 2005) at optical
and some NIR wavelengths towards 83 OB stars in the inner $2^\prime$
radius of R\,136. This work indicates that the extinction is variable in
the field and $R_V$ is typically in the range $4  - 5$. On the other
hand, these early-type massive stars do not necessarily sample the same
environment in which low-mass stars form (e.g. De Marchi, Panagia \&
Sabbi 2011), and due to their short life they cannot reveal the
long-term evolution of the dust grains in those regions. In other words,
massive stars only allow us to probe the most active star-forming
environments at the peak of the burst, but those conditions are not
characteristic of all objects in the field.

To overcome these limitations, we have developed a new method to
unambiguously determine the absolute value of the extinction in a
uniform way across a field such as that of 30\,Dor (De Marchi, Panagia
\& Girardi 2014; hereafter Paper\,I). Our method makes use of
multi-band  photometry of red giant stars belonging to the red clump
(RC). Other authors have used RC observations in the past to study the
reddening distribution and to derive reddening maps in the LMC (e.g.
Zaritsky 1999; Haschke et al. 2011; Tatton et al. 2013). However, all
these works have assumed an extinction law and therefore cannot derive
it independently. 

Applied to a typical field in the Magellanic Clouds, such as 30\,Dor,
our method offers many advantages: {\em (i)} all RC stars on which we
operate are at the same distance, to better than 1\,\%; {\em (ii)} they
have very similar intrinsic physical properties in all bands, within
$0.05$ mag for similar age and metallicity; {\em (iii)} our statistics
is very solid, with about 20 stars arcmin$^{\rm -2}$ or $\sim 150$
objects in a typical {\em Hubble Space Telescope} (HST) pointing; and
{\em (iv)} we derive a self-consistent absolute extinction curve over the
entire wavelength range covered by the photometry with no need for 
spectroscopic observations. This method can be easily extended to other
nearby galaxies.

In this work, we apply our method to the panchromatic observations of 
30\,Dor in the range $0.3 - 1.6\,\muup$m collected in 2009 with the
WFC\,3 instrument on board the HST (De Marchi et al. 2011a). The
structure of the paper is as follows. The observations are presented in
Section 2, while the selection of RC stars is discussed in Section 3.
Section 4 is devoted to deriving the absolute extinction towards RC
stars and the corresponding extinction law. In Section 5 we discuss the
reddening distribution in this field. A summary and our conclusions
follow in Section 6.

\section{Observations and data analysis}

\begin{figure*}
\centering
\resizebox{\hsize}{!}{\includegraphics[width=16cm]{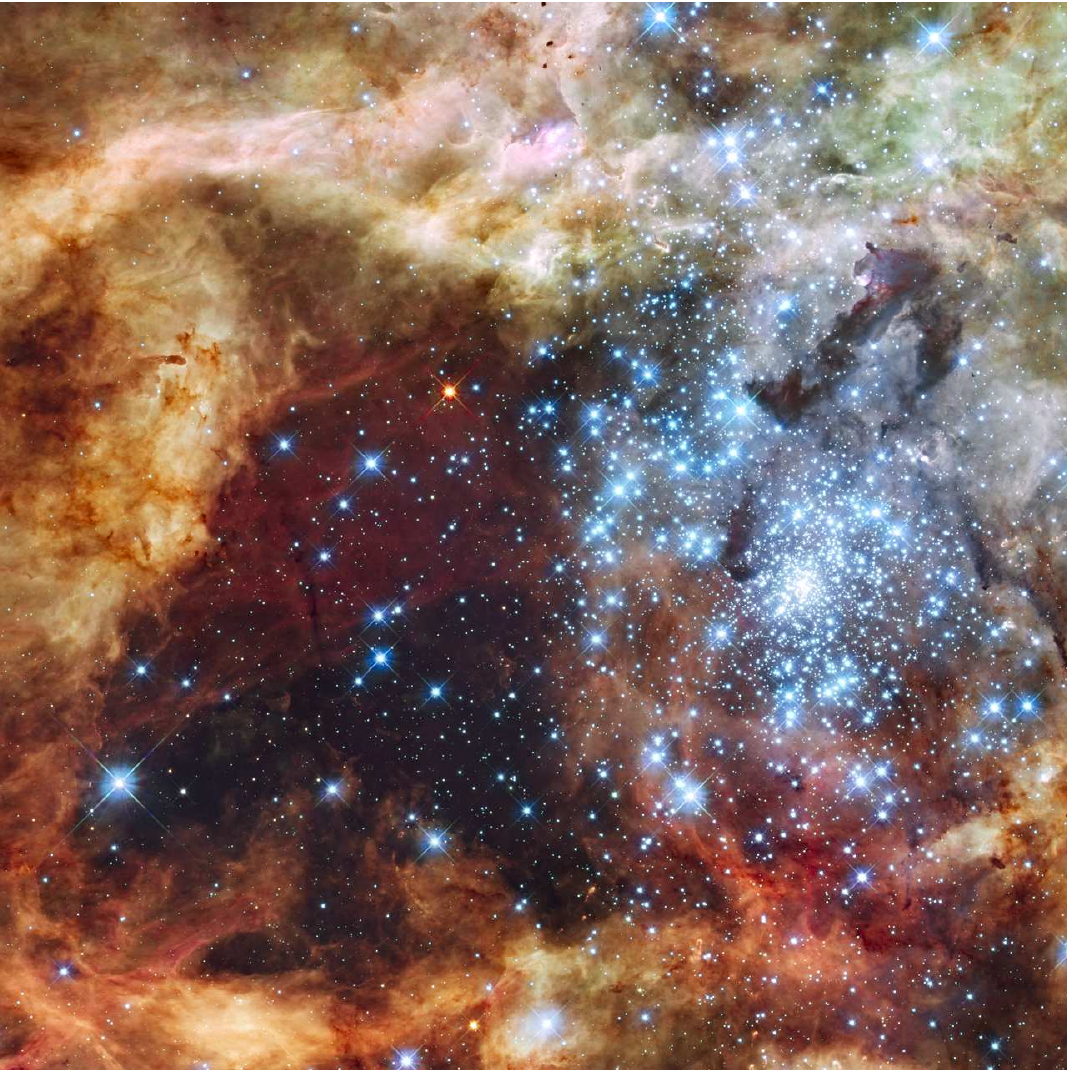}
                      \includegraphics[width=16cm]{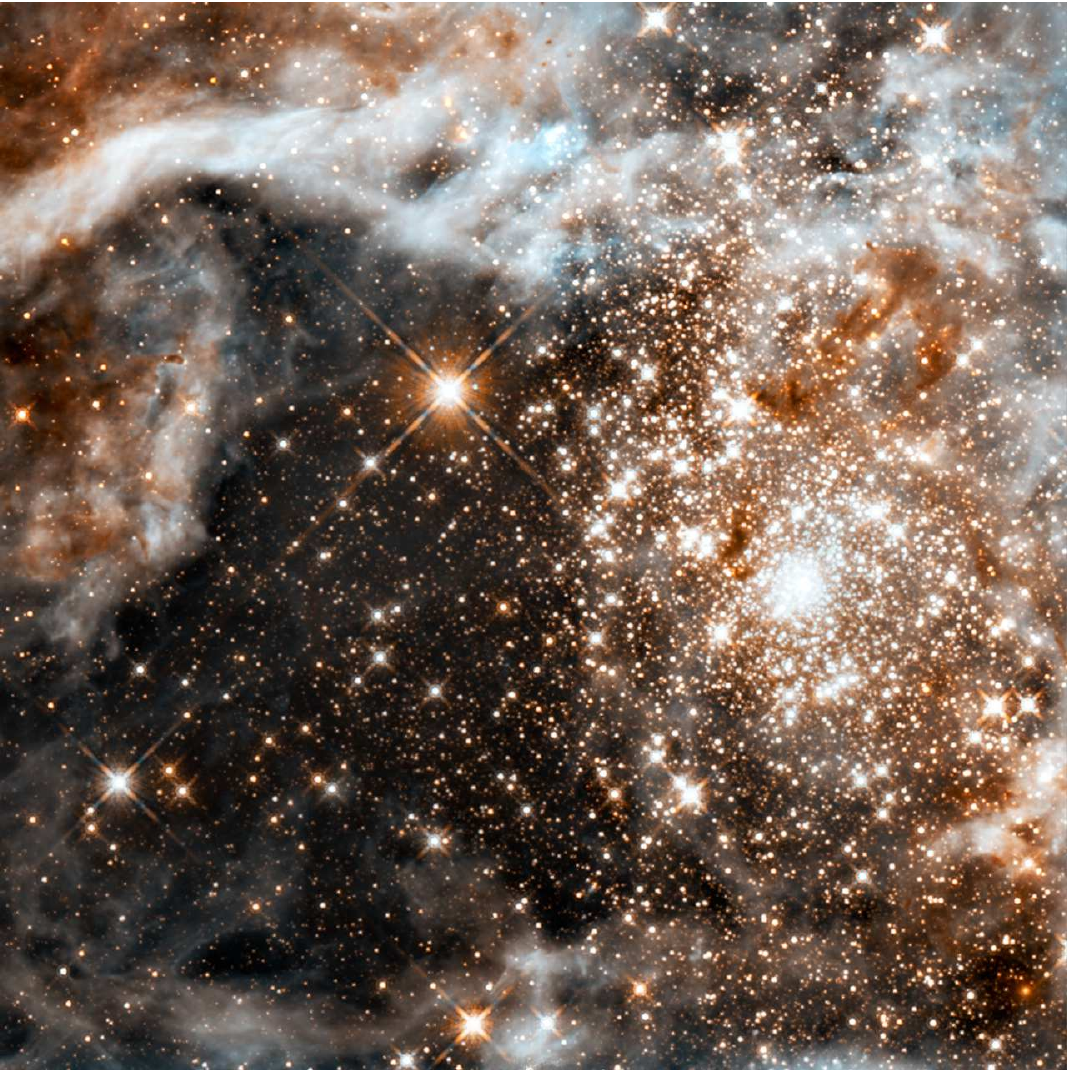}}
\caption{Optical (left) and NIR (right) colour-composite images of  a
region of $\sim 2^\prime \times 2^\prime$  at the centre of the field
covered by the observations. In the left panel the colours are as
follows: the blue channel is the average of F336W and F438W, the green
channel is F555W, the red channel F814W, and orange is provided by
F656N. In the right panel, the F110W and F160W filters are  assigned
respectively to complementary teal and orange. North is inclined 14 deg
to the left of the vertical and east is to its left.  (These images were
released by the Space Telescope Science Institute  on 2009 December 15
as part of News Release number STScI-2009-32.)}
\label{fig1}
\end{figure*}

The observations used in this work were obtained in 2009 October using
the WFC\,3 camera on board the HST, and are extensively described in De
Marchi et al. (2011a). The filters used, the number of exposures and
their total durations are listed in Table\,\ref{tab1}.  The observations
made use of an extensive dithering pattern, combining long and short
exposures, in order to improve the sampling of the telescope's point
spread function and increase the dynamic range of the photometry.

Figure\,\ref{fig1} shows a region of about $2^\prime \times 2^\prime$ 
at the centre of the field covered by the observations (the total area
covered spans $2\farcm7 \times 2\farcm7$). The bright cluster in the
field is R\,136. The panel on the left provides a colour-composite, with
the blue channel obtained by averaging the F336W and F439W filters, the
F555W filter serves as green channel, the F814W filter as red channel
and the F656N filter is used as orange. The panel on the right results
from the combination of the F110W and F160W filters, assigned
respectively to complementary teal and orange colours.

\begin{table}
\centering 
\caption{Number of exposures and cumulative exposure times in the 
various bands.}
\begin{tabular}{llcc} 
\hline
Filter & Band &  $N_{\rm exp}$  & $t_{\rm tot}$ \\
\hline  
F336W & ($U$) & 24 & $~\,8\,659$\,s\\
F438W & ($B$) & 16 & $~\,5\,174$\,s\\
F555W & ($V$) & 20 & $~\,6\,892$\,s\\
F656N & ($H\alpha$)& ~\,8 & $10\,805$\,s\\
F814W & ($I$) & 20 & $10\,700$\,s\\ 
F110W & ($J$) & ~\,9& $~\,1\,518$\,s\\
F160W & ($H$) & 12 & $~\,7\,816$\,s\\
\hline
\end{tabular}
\label{tab1}
\end{table}

\begin{figure*}
\centering
\resizebox{\hsize}{!}{\includegraphics[bb= 1 1 392 433]{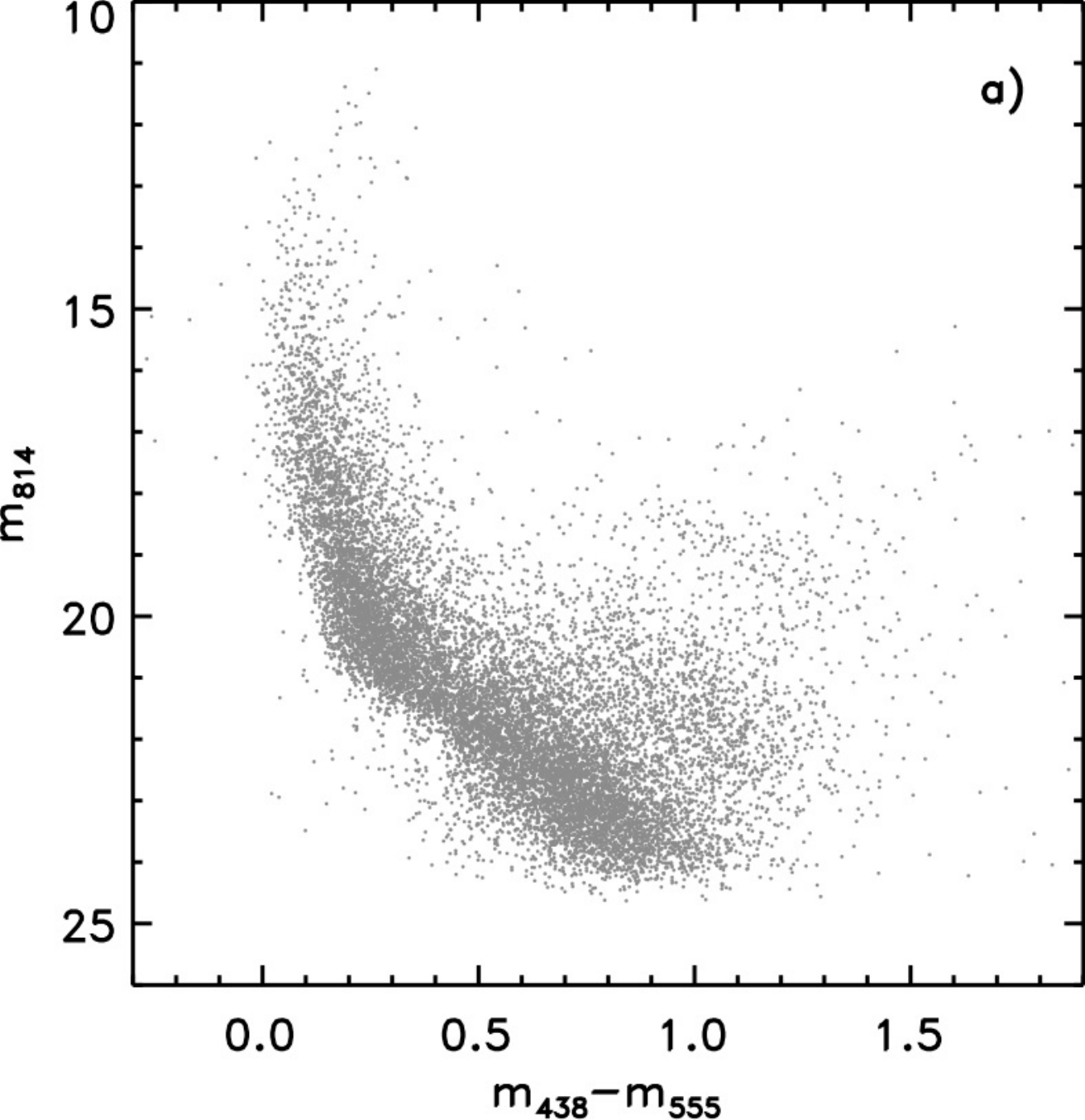}
                      \includegraphics[bb= -30 1 480 433]{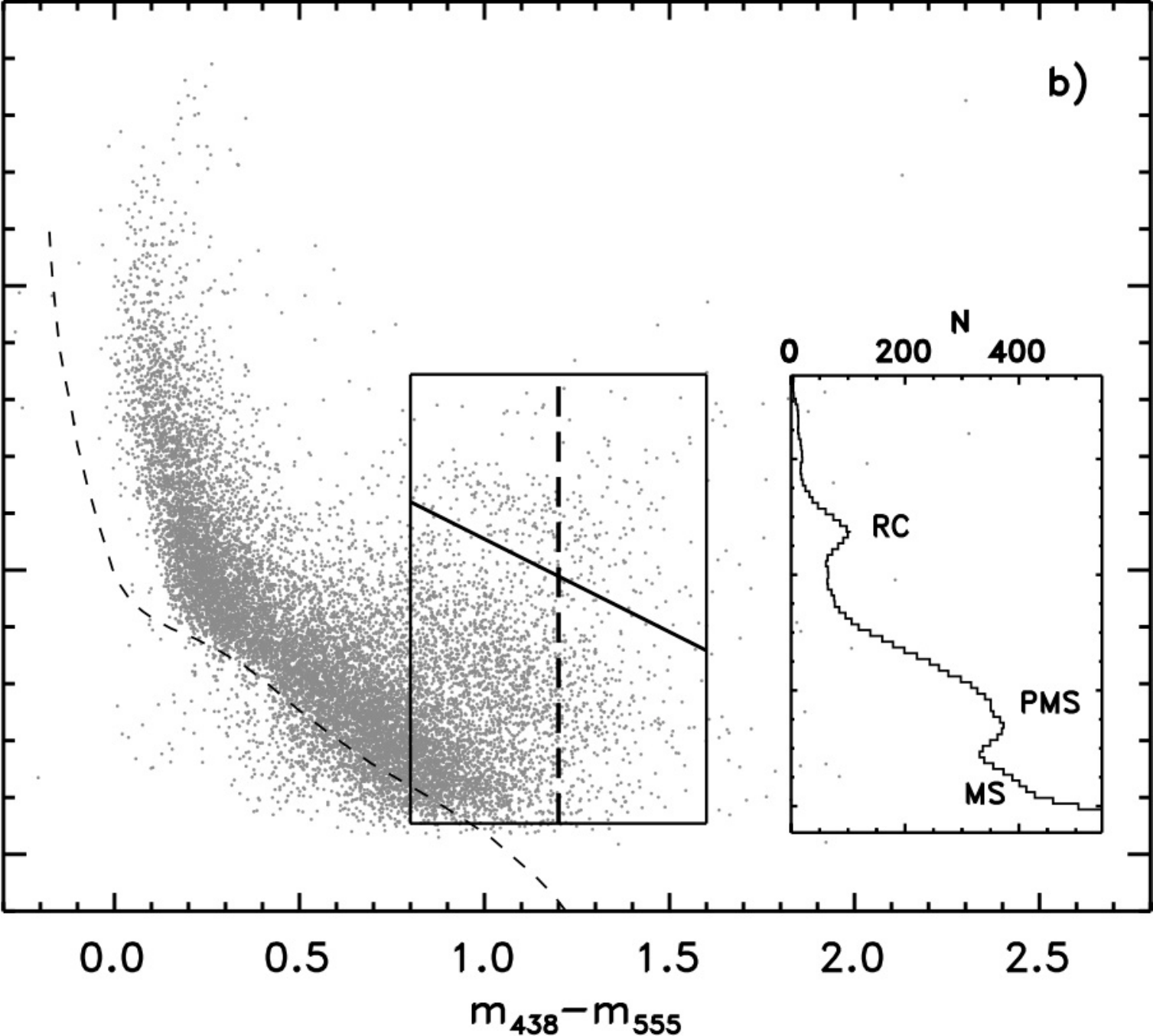}
                      \includegraphics[bb= 1 1 342 433]{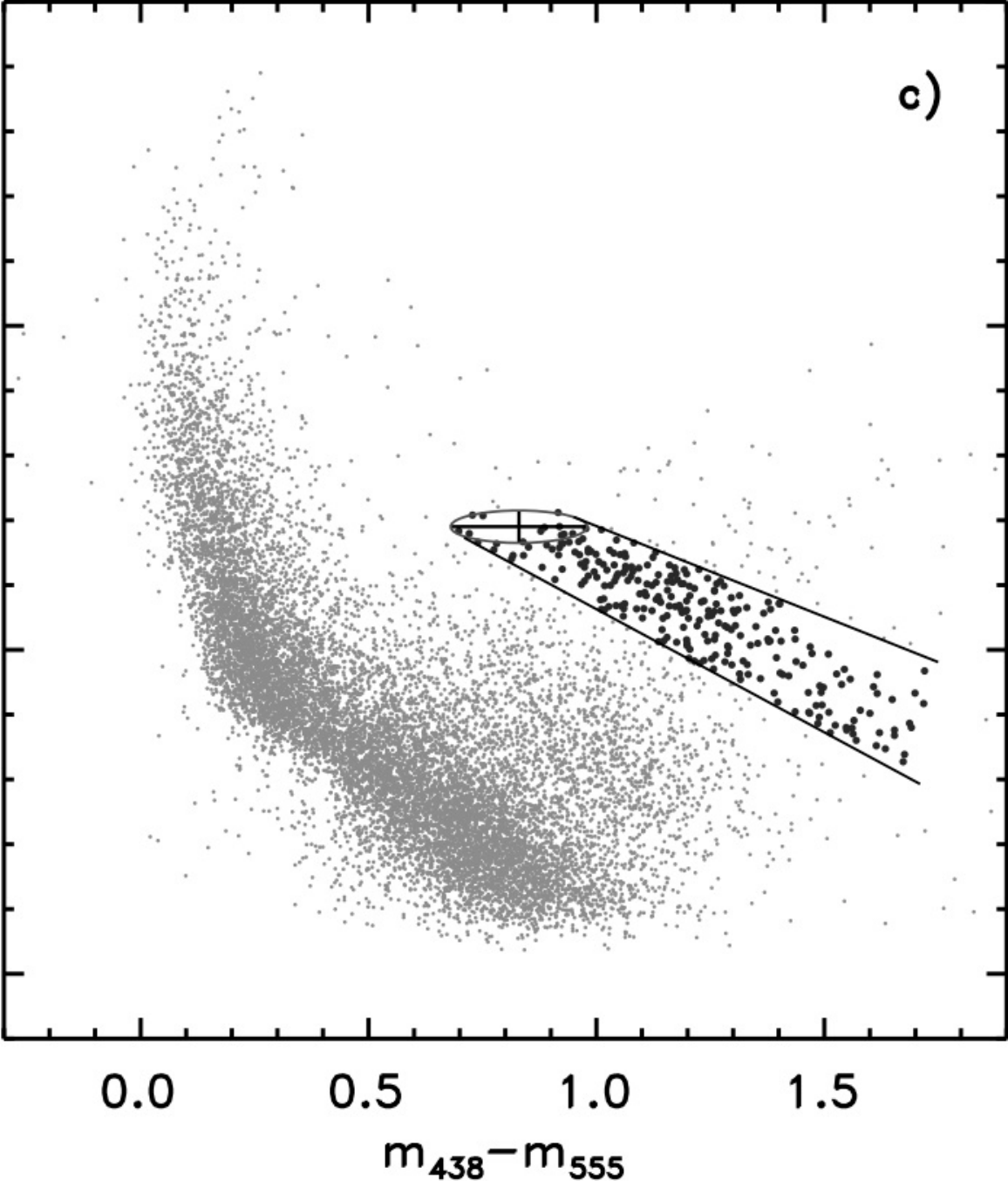}}
\caption{Panel {\bf a)}: As an example, we show the CMD in the $B$, $V$,
and $I$ bands. Only stars  with $\delta_4 < 0.1$ are included (the
typical photometric uncertainty is comparable with the size of the dots
in the figure). RC stars define a prominent sequence almost parallel to
the low-mass MS but well separated from it and from the PMS stars. Panel
{\bf b)}: We characterize the separation between RC and MS stars. The
apparent slope of the RC sequence is indicated by the tilted solid line
in the central box, while the ZAMS is indicated by the thin dashed
curve, from the models of Marigo et al. (2008), for the distance to the
LMC and already including intervening Galactic extinction. The inset on
the right-hand side shows  the number of objects contained in slices
parallel to the RC sequence as a function of magnitude. The density
peaks corresponding to the RC and PMS  stars are distinct and clearly
visible. Panel {\bf b)}: the initial selection of  candidate RC stars
includes all objects with $\delta_4 < 0.1$ located inside the
theoretical RC error ellipse or contained between the two tangents to
it. The slope of the tangents is that of the approximate reddening
vector with an added 15\,\% uncertainty. Candidate RC stars are
identified in the same way in all CMDs. }
\label{fig2}
\end{figure*}

Also the photometry is extensively discussed in De Marchi et al.
(2011a), to whom we refer the reader for further details. As an example,
we show in Figure\,\ref{fig2} the $I$ {\em vs} $B-V$ colour--magnitude
diagram (CMD), compared with the zero age main sequence (ZAMS) from the
models of Marigo et al. (2008; thin dashed line, extending up to
60\,\Msolar) for the specific WFC\,3 filters used here and a metallicity
of $Z=0.007$ as appropriate for R\,136 and the young LMC population in
general (e.g. Hill, Andrievsky \& Spite 1995; Geha et al. 1998). As for
the distance modulus, we have assumed $(m-M)_0 = 18.6$ (Panagia et al.
1991; Panagia 1999; Walborn \& Blades 1997). Only objects with combined
photometric uncertainty in the four optical broad bands smaller than
$0.1$\,mag are shown in Figure\,\ref{fig2}. The combined photometric
uncertainty $\delta_4$ is defined by Romaniello (1998) as:

\begin{equation}
\delta_4 = \sqrt{\frac{\delta^2_{336} + \delta^2_{438} +
                       \delta^2_{555} + \delta^2_{814}}{4}}
\label{eq1}
\end{equation}

\noindent 
where $\delta_{336}$, $\delta_{438}$, $\delta_{555}$, and $\delta_{814}$
are the uncertainties in each individual band.\footnote{The definition
given by Equation\,\ref{eq1} can be generalised for any combination of
bands. For instance, in Section\,5 we will also refer to the combined
photometric uncertainty $\delta_3$ in the $B$ (F438W), $V$ (F555W) and
$I$ (F814W) bands. For details on the photometric uncertainty in the
individual bands, see De Marchi et al. (2011a).} The ZAMS is also 
already reddened by the amount corresponding to the foreground Milky Way
(MW) absorption along the line of sight, which Fitzpatrick \& Savage
(1984) estimated to be $E(B-V)=0.07$ or $A_V = 0.22$. The comparison
shows that there is an additional extinction component in front of
30\,Dor, intrinsic to the LMC, which is known for a long time (e.g.
Fitzpatrick \& Savage 1984). Furthermore, a comparison of the width of the
upper main sequence (MS) with the photometric uncertainties quoted above
confirms that there is differential reddening in the field, as already
reported elsewhere (e.g. Hunter et al. 1995; De Marchi et al. 2011a).
The goal  of this work is to determine the amount and type of extinction
in this field and the corresponding properties of the ISM. As mentioned
in the Introduction, to this purpose we will use the RC objects present
in the field.

\section{Red clump stars}

{In order to determine the extinction curve in this field, as
discussed by De Marchi, Panagia \& Girardi (2014; Paper\,I), it is
necessary to identify the bona-fide RC stars and to measure their
displacement in the CMD from the location that they would occupy in the
absence of extinction. The procedure has three main steps, namely the
identification of the candidate RC stars, the removal of possible
outliers, and the determination of the slope of the reddening vector in
each set of bands. The steps are carried out separately in each of the
eight CMDs built with our set of bands as a function of the
$m_{438}-m_{555}$ and $m_{555}-m_{814}$ colours. However, since the NIR
observations do not cover the entire field, the primary selection of RC
objects is based on the four optical bands.}

\subsection{Identification of candidate RC stars}

We take as starting location the ``nominal RC,''
defined as the theoretical RC of stars of the lowest metallicity 
consistent with the observations. In Paper\,I we concluded that the
most  appropriate metallicity for the old stars ($> 1$\,Gyr) in the
field around the 30\,Dor region is $Z=0.004$. This choice  defines the
magnitude of the nominal RC, whereas for the colour we take the average
RC colours of stars in the range $1.4 - 3.0$\,Gyr, as in Paper\,I. 

The corresponding apparent magnitudes and associated $1\,\sigma$ spread
are shown in Table\,\ref{tab2}, while the colours in the most common
band combinations are $m_{438}-m_{555} = 0.83 \pm 0.06$,
$m_{555}-m_{814}= 1.06 \pm 0.04$. These magnitudes already include the
effects of the foreground MW extinction, {\em i.e.}  $E(B-V)=0.07$ or
$A_V=0.22$ as mentioned above (Fitzpatrick \& Savage 1984). 

\begin{table}
\centering 
\caption{Apparent magnitudes $m_{\rm RC}$ of the RC and corresponding 
$1\,\sigma$ spread in all bands, already including the effects of the
intervening MW extinction.}
\begin{tabular}{llcc} 
\hline
Filter & Band &  $m_{\rm RC}$ & $\sigma$\\
\hline 
F336W & $(U)$ & $20.34$ & $0.12$\\
F438W & $(B)$ & $19.98$ & $0.10$\\
F555W & $(V)$ & $19.16$ & $0.08$\\
F814W & $(I)$ & $18.10$ & $0.08$\\ 
F110W & $(J)$ & $17.72$ & $0.10$\\
F160W & $(H)$ & $17.15$ & $0.10$\\
\hline
\end{tabular}
\label{tab2}
\end{table}

{An inspection of the CMDs (see as an example Figure\,\ref{fig2}a) 
immediately reveals the candidate RC objects, as they define a prominent
sequence almost parallel to the MS but well separated from it and from
the pre-main sequence (PMS) stars. 

To confirm that the candidate RC stars are distinct from the MS objects,
we characterize their separation in a quantitative way, by calculating}
the number of stars inside slanted slices, parallel to one another and
progressively fainter. The slope of the slices is that defined by the
apparent slope of the extinguished RC sequence in the CMD and it is
shown by the tilted solid line in Figure\,\ref{fig2}b. In this example
the slices are $0.1$\,mag thick and are contained within the box shown
at the centre of the figure. The total number of objects in each slice
is shown by the histogram in the inset at the right-hand side of
Figure\,\ref{fig2}b, as a function of the magnitude. The vertical scale
is shifted in such a way that the peaks in the histogram are shown at
the magnitudes that correspond to the intersection of the reddening
vector (tilted solid line) with the median of the box (vertical dashed
line). The density peaks corresponding to the RC and PMS stars are 
{distinct and} clearly visible, and so is the sharply increasing ramp due
to MS stars.

The actual selection of candidate RC stars is done separately in each
CMD, i.e. those obtained by plotting the magnitudes in all bands as a
function of $m_{438}-m_{555}$ and $m_{555}-m_{814}$. After determining
the approximate direction of the reddening vector {(see the tilted
solid line in Figure\,\ref{fig2}b for the specific CMD shown there)}, we
plot the location of the theoretical RC and trace two lines tangent the
error ellipse defined by $2.5\,\sigma$. The slope of the tangents is
that of the approximate reddening vector with an added 15\,\%
uncertainty. {This is done separately for each CMD and  as} an
example we show in Figure\,\ref{fig2}c the case of the CMD in the
$m_{814}$ {\em vs.} $m_{438}-m_{555}$ bands (corresponding to $I$ {\em
vs.} $B-V$). In each CMD se take as candidate RC stars those inside the
region defined by the error ellipse and its tangents. These objects are
marked as thicker dots in Figure\,\ref{fig2}b.

We repeat this procedure in the CMDs resulting from all combinations of
bands as a function of the $m_{438}-m_{555}$ and $m_{555}-m_{814}$
colours, since those exist for all objects in the field, deriving in
this way eight sets of candidate RC objects. In the most conservative
approach, we retain as bona-fide RC stars only the objects that satisfy
this condition simultaneously in all optical bands, {\em i.e.,} a total
of 146 stars. We cannot impose the same conditions for stars observed
through the NIR bands, since those observations cover a more limited
area than the optical bands. Therefore, we take as candidate RC objects
in those bands all those that are present in the photometry and are
classified as RC stars in all other bands simultaneously. In total there
are 93 such objects.

\begin{figure*} \centering
\resizebox{!}{20cm}{\includegraphics{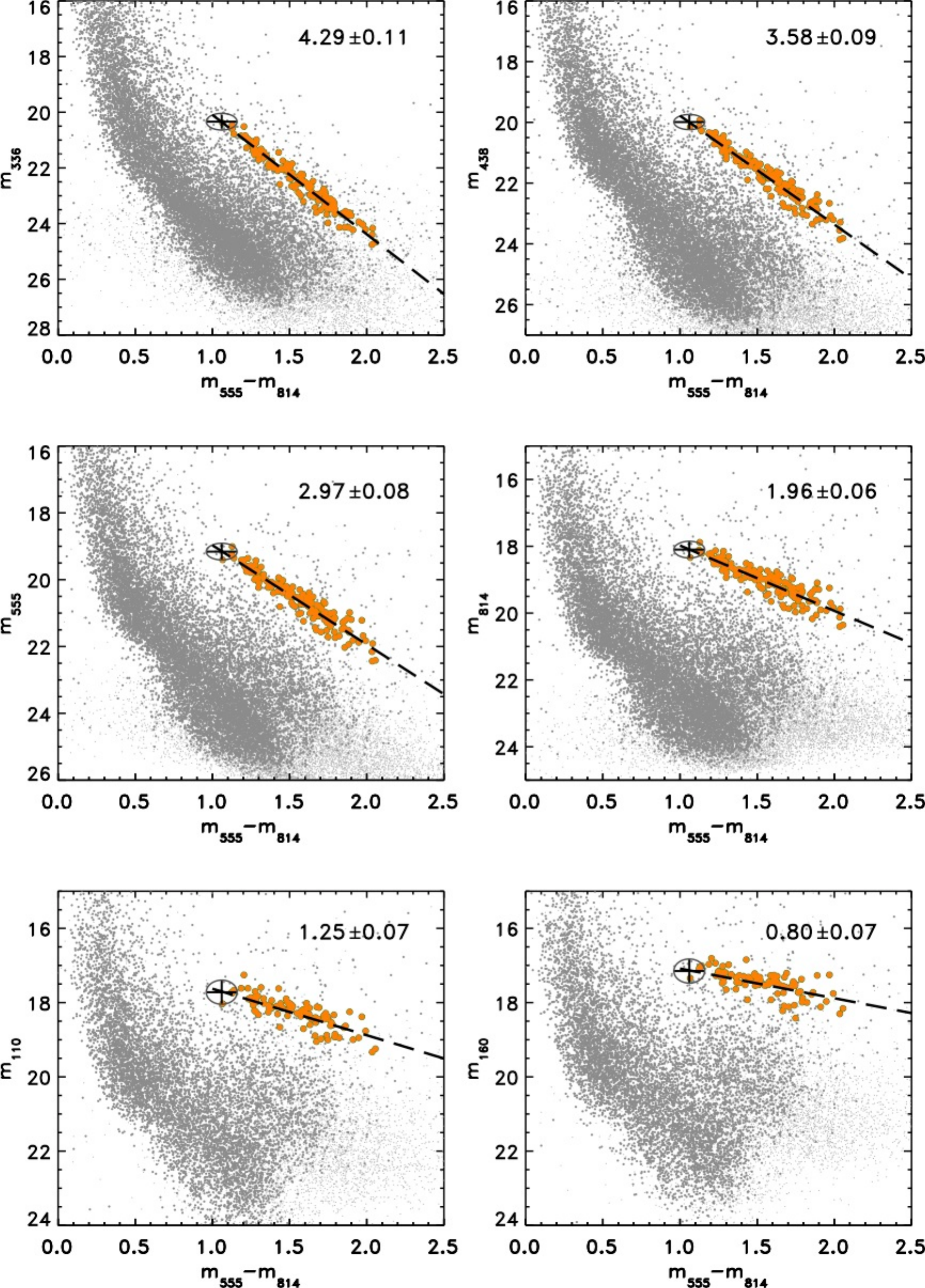}}
\caption{CMDs as a function of the $m_{555}-m_{814}$ colour. The  thick
orange dots are the the 140 bona-fide  RC stars (90 of which are also
are visible in the NIR bands). The nominal location of the RC is shown
by the ellipses in each graph and the best-fitting reddening vectors are
indicated by the dashed lines. The value of the parameter $R$ and its
uncertainty are given in each panel.} 
\label{fig3} 
\end{figure*}

\subsection{Removal of possible outliers}

{In order to guarantee that our RC candidate stars have physical
properties consistent with those of this class of objects, we } have
further excluded from this sample a total of 6 stars that have 
H$\alpha$ equivalent width $W_{\rm eq}(H\alpha) > 3$\,\AA, as they could
be PMS stars (three of them also fall in the area covered by the NIR
observations). Objects with H$\alpha$ excess emission are easily
identified following the method described by De Marchi, Panagia \&
Romaniello (2010) and De Marchi et al. (2010b), to which we refer the
reader for more details. 

As shown in De Marchi et al. (2011a), there is a large population of PMS
stars in this field and some of them have colours and magnitudes that
place them in the CMD region occupied by extinguished RC objects. It is
well known that also chromospheric activity is a likely source of
H$\alpha$ emission in stars of low mass such as those in the RC phase.
However, as discussed in Paper\,I, White \& Basri (2003) indicate
specific thresholds above which stars with H$\alpha$ excess emission
should be considered classical T Tauri stars, instead of just
chromospherically active objects. 

For stars of spectral type K0 -- K5, White \& Basri (2003) recommend
thresholds of 3\,\AA, and values larger than 10\,\AA\ for K7 -- M2.5
stars. Since these interlopers, like the RC stars, are affected by
reddening, we cannot determine their actual spectral type. Thus, we make
the most conservative assumption that they are all of spectral type
earlier than K0 and exclude from the bona-fide sample of RC stars a
total of 6 sources that have $W_{\rm eq}(H\alpha) > 3$\,\AA, of which 
3 also fall in the area covered by the NIR observations. The final
count of bona-fide RC stars includes 140 objects detected in all optical 
bands, of which 90 have also been measured through the NIR filters.
The bona-fide RC stars are shown as thick orange dots in the CMDs of
Figure\,\ref{fig3}.

\subsection{Deriving the slope of the reddening vectors}

As discussed in Paper\,I, in each CMD we derive the best linear fit to 
the distribution of the bona-fide objects taking into account: {\em 1)}
the uncertainties on their magnitudes and colours; {\em 2)} the
uncertainties on the ellipse defining the model RC in each individual
CMD; {\em 3)} an equal number (140 or 90 depending on the bands) of 
synthetic un-reddened RC stars with a Gaussian distribution inside the 
ellipses. The latter step guarantees that the best linear fit passes
through the centre of the nominal RC. The reddening slopes derived in
this way correspond to the value of the ratio $R$ between absolute and
selective extinction in the specific bands of our observations. The
values of the ratio $R$ and the corresponding uncertainties are listed
in Table\,\ref{tab3}. The reddening slopes are also shown by the thick 
dashed line in the CMDs of Figure\,\ref{fig3}, where all stars in our
photometric catalogue are shown. The thick orange dots are the
bona-fide  RC stars, while the light- and dark-grey smaller dots are for
stars with, respectively, $\delta_4$ photometric uncertainty larger or
smaller than $0.1$\,mag. The values of $R$ are also indicated in each
panel.

\begin{table}
\centering 
\caption{Measured values of the ratio $R$ between absolute ($A$) and 
selective ($E$) extinction in the specific bands of our observations, 
with corresponding uncertainties. The effective wavelength ($\lambda$)
and wave number ($1/\lambda$) of each band are also indicated.}
\begin{tabular}{cccc} 
\hline
Band combination & $R$ & $\lambda$  & $1/\lambda$\\
 & & [\AA] & [$\muup$m$^{-1}$]\\
\hline
$A_{336}/E(m_{438}-m_{555})$ & $6.81 \pm 0.34$ & ~3\,359 & $2.98$ \\
$A_{438}/E(m_{438}-m_{555})$ & $5.73 \pm 0.29$ & ~4\,332 & $2.31$ \\
$A_{555}/E(m_{438}-m_{555})$ & $4.79 \pm 0.25$ & ~5\,339 & $1.87$ \\
$A_{814}/E(m_{438}-m_{555})$ & $3.24 \pm 0.18$ & ~8\,060 & $1.24$ \\
$A_{110}/E(m_{438}-m_{555})$ & $2.20 \pm 0.18$ & 11\,608 & $0.86$ \\
$A_{160}/E(m_{438}-m_{555})$ & $1.52 \pm 0.16$ & 15\,387 & $0.65$ \\
\hline
$A_{336}/E(m_{555}-m_{814})$ & $4.29 \pm 0.11$ & ~3\,359 & $2.98$ \\
$A_{438}/E(m_{555}-m_{814})$ & $3.58 \pm 0.09$ & ~4\,332 & $2.31$ \\
$A_{555}/E(m_{555}-m_{814})$ & $2.97 \pm 0.08$ & ~5\,339 & $1.87$ \\
$A_{814}/E(m_{555}-m_{814})$ & $1.96 \pm 0.06$ & ~8\,060 & $1.24$ \\
$A_{110}/E(m_{555}-m_{814})$ & $1.25 \pm 0.07$ & 11\,608 & $0.86$ \\
$A_{160}/E(m_{555}-m_{814})$ & $0.80 \pm 0.07$ & 15\,387 & $0.65$ \\
\hline      
\end{tabular}
\vspace{0.5cm}
\label{tab3}
\end{table} 

\section{Extinction law}

It is customary to express the extinction law in the form of the ratio

\begin{equation}
R_\lambda \equiv \frac{A_\lambda}{E(B-V)},
\label{eq2}
\end{equation}

\noindent  
where $A_\lambda$ is the extinction in the specific band and $E(B-V)$
the colour excess in the canonical Johnson $B$ and $V$ bands. The values
of $R_\lambda$ listed in Table\,\ref{tab3} do not exactly correspond to
Equation\,\ref{eq2}, since the colour excess is the one measured in the 
specific WFC\,3 bands used in making these observations. However, we can
easily derive the values of $R_\lambda$ through spline interpolation, as
described in Paper\,I, and can translate in this way the
values of Table\,\ref{tab3} into the corresponding values as a function
of $E(B-V)$ in the traditional Johnson bands. 

The $R_\lambda$ values obtained in this way are shown graphically in
Figure\,\ref{fig4}, where squares and triangles correspond to values
obtained originally from measurements as a function of the colour
indices $m_{438}-m_{555}$ and $m_{555}-m_{814}$, respectively. The two
sets of values are in excellent agreement with one other within their
uncertainties. The thick solid line shows a spline interpolation through
the $R_\lambda$ values and takes into account the  smaller uncertainties
in $m_{555}-m_{814}$. The interpolated $R_\lambda$ values in the
classical Johnson--Cousin bands, at the wavelengths marked by the
vertical dotted lines in the figure, are listed in Table\,\ref{tab4}
(note that the value for the $K$ band is actually an extrapolation and,
therefore, is indicated in italics in the table). For comparison, in
Figure\,\ref{fig4} we also show  the $R_\lambda$ values from Paper\,I,
measured in a field $\sim 6\arcmin$ SW of R\,136, and the corresponding
spline interpolation (long-dashed line), as well as the canonical
Galactic extinction law, taken from the work of Fitzpatrick \& Massa
(1990) for $R_V=3.1$ (short dashed line).

\subsection{Comparison with previous works}

The shape of the extinction curve that we obtain is not reproduced by
the traditional parameterizations based on the value $R_V$  ({\em e.g.,}
Cardelli et al. 1989; Fitzpatrick \& Massa 1990) because they are unable
to reproduce the entire wavelength range covered by our observations
with a single value of the parameter $R_V$. 

\begin{figure*}
\centering
\resizebox{\hsize}{!}{\includegraphics[width=14cm]{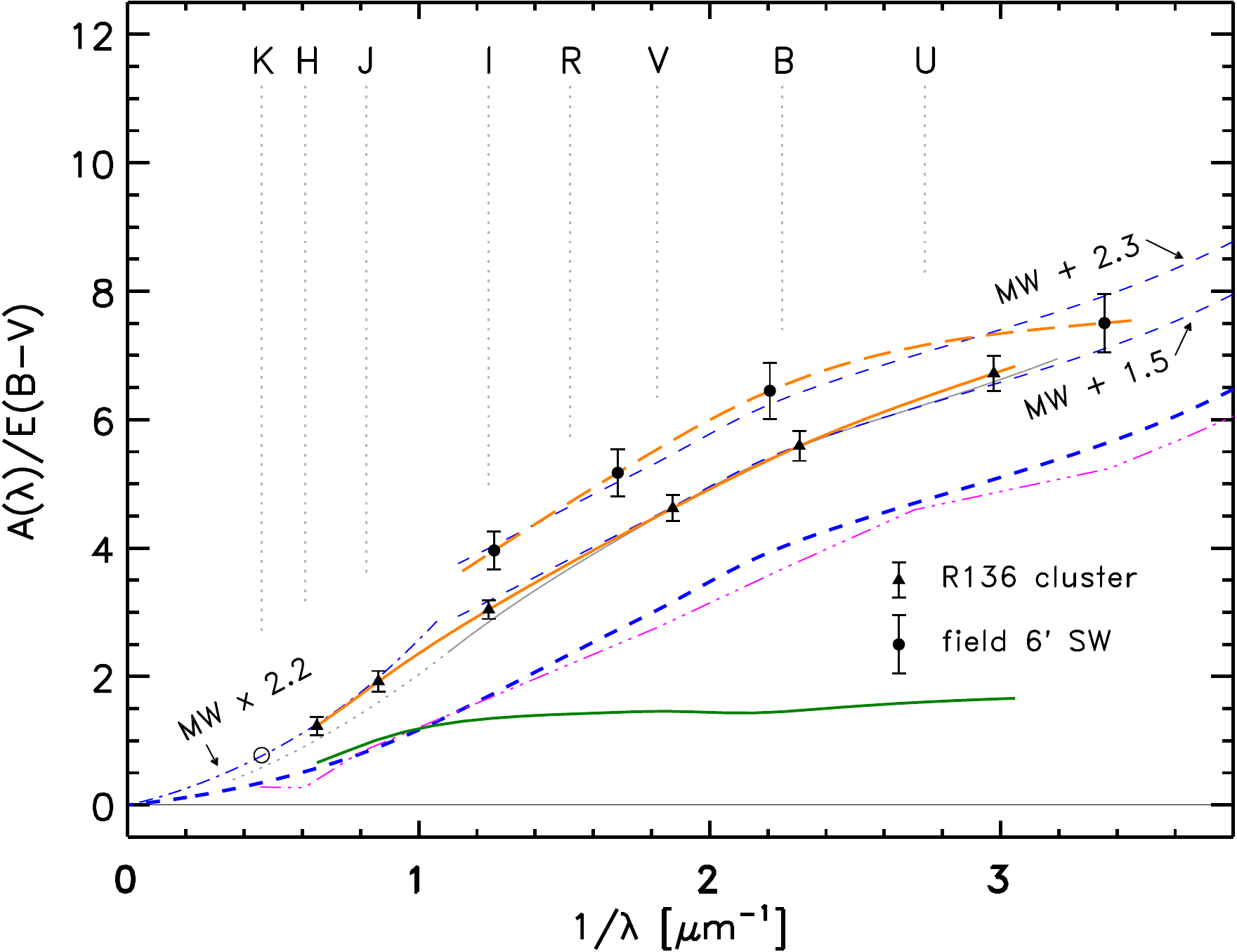}}
\caption{Extinction law. The squares and triangles show the measurements
derived in the core of 30\,Dor from the CMDs as a function of,
respectively, $m_{438}-m_{555}$ and $m_{555}-m_{814}$ colours (the error
bars for the $m_{555}-m_{814}$ data are often smaller than the symbols).
The dots correspond to the measurements in the region $6\arcmin$ SW of
the centre studied in Paper\,I. The solid and long-dashed lines are a
spline interpolation through those values. The short-dashed line
indicates the canonical Galactic extinction law, taken from Fitzpatrick
\& Massa (1999) for $R_V=3.1$. The thin dashed lines show the same law,
shortwards of $\sim 1\,\muup$m shifted vertically by $1.5$ and $2.3$.
The dot-dashed line is the Galactic extinction law, longwards $\sim
1\,\muup$m, multiplied by $2.2$ to fit the measured values in the $J$
and $H$ bands. The green solid line is the difference between the
Galactic extinction law (thick dashed line) and that of the R\,136
cluster (orange solid line).  The thin solid line, extended by a dotted
line longwards of 9\,000\AA, is the extinction law of Ma\'{\i}z
Apell\'aniz et al. (2014) for $R_V=4.4$. The thin triple-dot-dashed line
is the extinction curve of Gordon et al. (2003) for a field around
30\,Dor.}
\label{fig4}
\end{figure*}

As an example, we show in Figure\,\ref{fig4} with a thin grey line the
extinction law recently derived for this specific field by
Ma\'{\i}z Apell\'aniz et al. (2014). These authors have used
spectroscopy and NIR photometry from the  VLT-FLAMES Tarantula Survey
(Evans et al. 2011) and the same HST optical photometry discussed in
this work (De Marchi et al. 2011a) for a selected sample of 83 stars of
spectral types O and B (see Section\,5). Using the  Bayesian code
CHORIZOS (Ma\'{\i}z Apell\'aniz 2004), they have fitted the available
photometry to a family of synthetic spectral energy distributions, with
different assumptions on the stellar parameters and the shape of the
extinction law. They conclude that the best family of extinction laws is
a slightly modified version of that of Cardelli et al. (1989), since the
latter does not provide a good fit to the observations in the $U$ band
when $R_V$ is large. At longer optical wavelengths this new family of
extinction laws does not deviate from Cardelli et al.'s and Ma\'{\i}z
Apell\'aniz et al. (2014) assumed it to be  exactly that in the NIR. 

\begin{figure}
\centering
\resizebox{\hsize}{!}{\includegraphics[width=16cm]{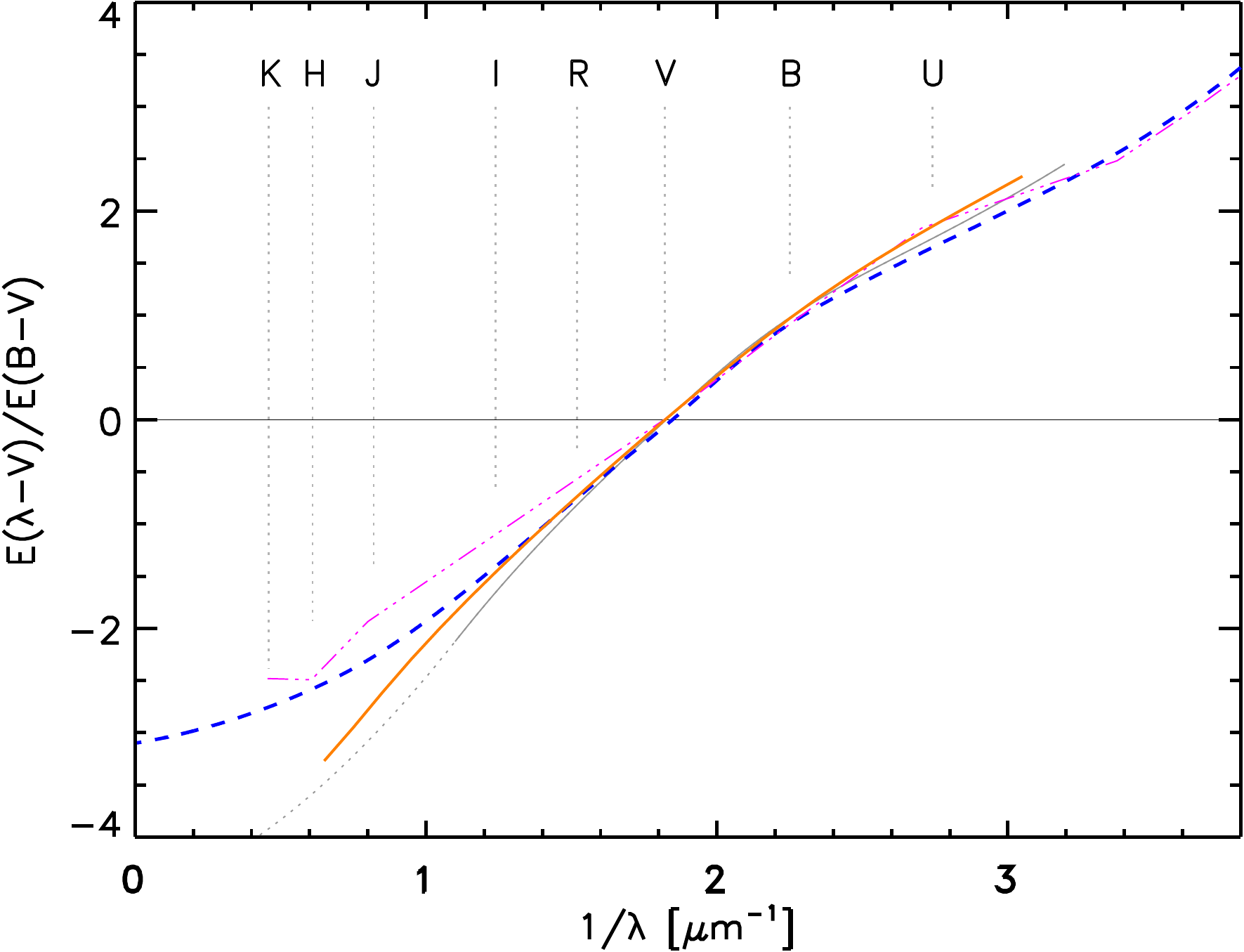}}
\caption{A subset of the extinction curves from Figure\,\ref{fig4} is
shown here, with the same line types, in units of the colour excess,
i.e. $E(\lambda-V)/E(B-V)$. This figure can make it easier to compare 
our results with previous works. Even when all curves are renormalized 
in the $V$ band, as in this figure, a marked difference remains at
$\lambda > 7\,000$\,\AA\ between  our measurements and the extinction
curve of Gordon et al. (2003; triple-dot-dashed line) longwards of the
$R$ band. }
\label{fig5}
\end{figure}

{Nor are our observations compatible with the extinction curve
derived by Gordon et al. (2003) from 8 sight lines around 30\,Dor. As
mentioned in the Introduction, these stars are outside the clusters at a
median distance of $\sim 20\arcmin$ from its centre. The extinction
curve of Gordon et al. (2003) is shown as a triple-dot-dashed line in
Figure\,\ref{fig4} and at optical wavelengths it approaches the Galactic
extinction law (short dashed line). Hence, although this extinction
curve cannot be considered representative of 30\,Dor, it provides a good
reference for the surrounding regions. The significant difference
between our measurements and the extinction curve of Gordon et al.
(2003), and at NIR wavelength also that of Ma\'{\i}z Apell\'aniz et al.
(2014), is also visible in Figure\,\ref{fig5}. A subset of the curves of
Figure\,\ref{fig4} are shown here in units of $E(\lambda-V)/E(B-V)
\equiv (A_\lambda-A_V)/E(B-V)  \equiv R_\lambda-R_V$, in such a way that
they are all normalised in the $V$ band. It is important to understand
that this type of graph only shows the {\em relative} differences
between the curves, but Figure\,\ref{fig4} reveals that there are also
important differences in the absolute value of the extinction, hence in
the value of $R_V$.}

\begin{table}
\centering 
\caption{Interpolated values of $R_\lambda$ for the most common bands.
The table also gives the effective  wavelengths ($\lambda$) and wave
numbers ($1/\lambda$) of the filters, the value of $R_\lambda^{MW}$ for
the canonical extinction law in the diffuse Galactic ISM, and the
difference between the latter and our measurements. All values are given
for the specific monochromatic effective wavelength as indicated, without
considering the width of the filters. Values in the $K$ band are
indicated in italics as they are extrapolated.}
\begin{tabular}{cccccc} 
\hline
Band & $\lambda$  & $1/\lambda$  &  $R_\lambda$ &
$R_\lambda^{MW}$ & $R_\lambda - R_\lambda^{MW}$\\
 & [\AA] & [$\muup$m$^{-1}$] & & & \\
\hline
$U$ & ~3\,650 &  $2.74$ & $6.35\pm0.23$ & $4.75$ & $1.60$ \\
$B$ & ~4\,450 &  $2.25$ & $5.48\pm0.20$ & $4.02$ & $1.46$ \\
$V$ & ~5\,510 &  $1.82$ & $4.48\pm0.17$ & $3.02$ & $1.46$ \\
$R$ & ~6\,580 &  $1.52$ & $3.77\pm0.15$ & $2.35$ & $1.42$ \\
$I$ & ~8\,060 &  $1.24$ & $3.04\pm0.13$ & $1.67$ & $1.34$ \\
$J$ & 12\,200 &  $0.82$ & $1.78\pm0.14$ & $0.83$ & $0.95$ \\
$H$ & 16\,300 &  $0.61$ & $1.12\pm0.12$ & $0.53$ & $0.59$ \\
$K$ & 21\,900 &  $0.46$ & $\mathit{0.78\pm0.09}$ & $0.34$ & 
  $\mathit{0.42}$ \\
\hline      
\end{tabular}
\vspace{0.5cm}
\label{tab4}
\end{table}

For the 83 OB stars in their sample, Ma\'{\i}z Apell\'aniz et al. (2014)
obtain an apparently wide spread of $R_V$ values, namely  $4.4 \pm 0.7$,
although the 50 stars with the smallest uncertainty on $R_V$ ($0.2$ or
less) indicate a smaller spread, namely $\pm 0.4$. The thin solid line
in Figure\,\ref{fig4} corresponds to $R_V=4.4$ and provides a good fit
to our optical data. 

{The very good agreement between the $R_V$ values measured by us and
by Ma\'{\i}z Apell\'aniz et al. (2014) is noteworthy, because they are
derived from stars of different types, namely RGs in our case and OB
stars in theirs. In principle, when the extinction is obtained from $B$
and $V$ photometry alone, the colour excess $E(B-V)$ corresponding to a
given value of $A_V$, and hence the derived value of $R_V$, will depend
on the intrinsic colour of the stars used for the measurements. McCall
(2004) has extensively discussed and quantified this effect, showing
that the $R_V$ value obtained from giants with $(B-V)_0 \simeq 1$ like
our RC stars could be about $\sim 10\,\%$ larger than that estimated
from hot stars with $(B-V)_0 \simeq 0$. This happens because the
effective wavelengths of both bands become redder, but more in $B$ than
in $V$ and hence  the extinction in $B$ is reduced more than in $V$. In
this way $E(B-V)$ is lowered with respect to $A_V$ and $R_V$ appears
higher. However, the importance of these effects is much lower if the
$V-I$ colour is used instead. As discussed in Section\,4, our extinction
law $R_\lambda$ is primarily based on those bands, owing to the smaller
photometric uncertainties in $m_{\rm 555} - m_{\rm 814}$ (see also
Table\,\ref{tab3}). Indeed, the excellent agreement between our $R_V$
value and that obtained from spectrophotometry by Ma\'{\i}z Apell\'aniz
et al. (2014) confirms that there are no systematic differences in the
optical domain.  

Interestingly, longwards of $\sim 9\,000$\,\AA\, a systematic deviation
between our curve and that of Ma\'{\i}z Apell\'aniz et al.'s (2014)} is
observed. A similar deviation is seen if we compare our extinction law
with that of Fitzpatrick \& Massa (1990) and Cardelli et al. (1989),
which is the functional form that Ma\'{\i}z Apell\'aniz et al. (2014)
have assumed in the NIR. However, they warn the reader that in the
NIR range alternatives to the Cardelli's law cannot be excluded, given
the small values of the NIR extinction of the stars in the sample. For
that reason, in Figure\,\ref{fig4} the curve of Ma\'{\i}z Apell\'aniz et
al. (2014) in shown with a dotted line in this spectral range. We
believe that the considerable complexity of the extinction phenomenon
(many different materials, many different sizes, etc.) makes it
excessively hard and perhaps even futile to define a simple parametric
curve to describe the observations accurately. Therefore, instead of
attempting a parametric fit to the extinction curve that we have
measured, we concentrate on the implications of its shape for the
physical properties of the grains.

\subsection{Extinction and properties of the grains}

With $R_V=4.5 \pm 0.2$, the extinction law that we measure around R\,136
is considerably flatter (i.e. less steep in logarithmic terms) than the 
extinction law of the Galactic ISM shown by the short dashed line. The
same is true for the extinction law in the field $6\arcmin$ SW, where
$R_V=5.6 \pm 0.3$ (Paper\,I). In linear terms, the extinction law in
these regions is actually almost exactly parallel to the Galactic curve,
as the thin dashed lines show. The latter are simply the portion of the
Galactic law shortwards of $1\,\muup$m shifted vertically by an offset
of $1.5$ and $2.3$, respectively, and provide a surprisingly good fit to
our observations inside R\,136 and in the neighbouring field.  Note
that also the optical portion of the extinction curve of Gordon et al.
(2003) is parallel to our measurements, and if shifted vertically by an 
offset of $1.6$ and $2.4$, respectively, it provides a good fit to our
observations. 

The green solid line shown in Figure\,\ref{fig4} represents the
difference between the extinction law in R\,136 and that in the Galaxy,
that we take as a template. The difference is indeed remarkably flat
shortwards of $\sim 1\,\muup$m (see also Table\,\ref{tab4}). This very
fact indicates that the extinction law in 30\,Dor is of a type similar
to that of the MW (and of the regions around the Tarantula nebula),  but
that there is an additional component, whose contribution is ``grey'' in
the optical, where it does not depend on wavelength. This means that
there must be a larger fraction of large grains, compared to that of the
diffuse Galactic ISM and the surroundings of 30\,Dor.

{At wavelengths longwards of $1\,\muup$m, the extinction law in 30\,Dor
falls off as $\sim \lambda^{-1.5}$, virtually indistinguishable from the
observed behaviour of the MW extinction law (Cardelli et al. 1989),
thereby indicating that the nature of the grains is the same.} Indeed,
the dot-dashed line shown in Figure\,\ref{fig4} is simply the portion of
the Galactic extinction law longwards of $1\,\muup$m multiplied by a
factor of $2.2$. We see that it fits the observations in the $J$ and $H$
bands quite well. 

A detailed modeling of the extinction curve is beyond the scope of this
work and will be deferred to a future paper. However, there are simple
considerations that can provide valuable insights into the properties of
the additional dust component that is present in the 30\,Dor nebula. 

It is well known ({\em e.g.,} van de Hulst 1957; Greenberg 1968; Draine 
\& Lee 1984) that at wavelengths short enough the extinction cross
section of a grain tends asymptotically to its geometric cross section
$\sigma_{\rm geom} = \pi \, a^2$, where $a$ denotes the grain radius. At
longer wavelengths the extinction is essentially pure absorption and the
cross section is smaller than $\sigma_{\rm geom}$, being proportional to
$\sigma_{\rm geom} \times 2\,\pi\,a / \lambda$. Conveniently
enough, the transition occurs approximately at $\lambda_0 \sim
2\,\pi\,a$. Thus, for a fixed grain size, one would expect a sort of a
step function behaviour with the transition occurring abruptly around
$\lambda_0$.

On the other hand, the Galactic extinction law is seen to increase
steadily with the wave number $\lambda^{-1}$, over a wide range of
wavelengths, say, from $\lambda \sim 5\,\muup$m or $\lambda^{-1} =
0.2\,\muup$m$^{-1}$ to $\lambda \sim 0.1\,\muup$m or $\lambda^{-1} =
10\,\muup$m$^{-1}$. This is the reason why to account for the MW
extinction law a grain size distribution as been invoked, with an
approximate form of a power-law $f(a) \propto a^{-3.5}$ within the
interval from $a_{\rm max} \sim 0.2\,\muup$m and $a_{\rm min} \sim 
0.01\,\muup$m (Mathis, Rumpl \& Nordsieck 1977; Draine \& Lee 1984).
With such a law one can reproduce the observations quite well,
essentially as a sum of step functions in which the large grain
extinction dominates at the longer wavelengths and progressively smaller
grains account for the increasing extinction at shorter wavelengths.

{It is worth noting that, for a size distribution $f(a) \propto
a^{-\beta}$ with $\beta < 4$, at wavelengths longer than $2\,\pi\,a_{\rm
max}$ the extinction is dominated by the largest grains and is
proportional to the total mass in grains.} As shown in
Figure\,\ref{fig4}, the 30\,Dor nebula extinction law can be represented
by the sum of a standard MW extinction curve plus one produced by an
additional component of large grains (solid brown curve). The fact that
in the NIR the total extinction towards R\,136 is about twice as high as
in the MW implies that the total mass in large grains is about twice as
high as in the MW. This conclusion should be regarded in relative terms,
in the sense that the relative abundance of the large grains must be a
factor of $2.2$ higher.

{Large $R$ values are typical of dense, still relatively quiescent
clouds, that have not yet undergone extensive star formation. An example
is the Orion nebula, where $R_V \simeq 5$ (e.g. Johnson 1967; Lee 1968).
Like in 30\,Dor, the dust in Orion is dense-cloud dust that is now
being  exposed to hot young stars and its extinction curve is dominated
by large grains (e.g. Cardelli \& Clayton 1988).} A higher relative
abundance of large grains may be obtained in at least three different
ways: {\em i)} selective destruction of small grains so as to skew the
size distribution in favour of large grains; {\em ii)} selective
condensation of material on the surface of small grains, making them
effectively bigger; {\em iii)} selective injection of ``fresh'' large
grains into the MW mix.

The first option may be discounted because it would imply an UV
extinction law that is flatter than that in the MW, whereas in fact a
steep rise in the UV has been measured towards stars of the Magellanic
Clouds, as compared to MW objects ({\em e.g.} Fitzpatrick 1998). The
second option would create the same problem as the first one and, in
addition, it would require an efficient condensation process to operate
on the surfaces of the grains.

The third scenario looks more likely, because adding new large grains
could easily account for the presence of the new component of the
extinction curve, without conflicting against any existing evidence at
other wavelengths, most notably the UV. Furthermore, injection of large
grains is an exciting possibility opened by recent {\em Herschel} and
{\em ALMA} observations of SN\,1987A (Matsuura et al. 2011; Indebetouw
et al. 2014). These observations suggest that a substantial amount
($> 0.4$\,\Msolar) of large grains ($> 0.1\,\muup$m)  were created in
the ejecta of the supernova. That large grains are indeed formed in SNe 
has been recently established for SN\,2010jl (Gall et al. 2014). Thus 
if one assumes that a comparable
output follows {\em all} Type\,II supernova explosions, it would be easy
to understand  why regions of recent star formation may display the
presence of a higher content of large grains. Within this scenario, one
would expect that in regions where star formation was active in the
recent past, say, $\sim 50$\,Myr (i.e. the lifetime of 8\,\Msolar\,
stars that represent the lower mass limit of supernova type II
progenitors), the excess of large grains should be highest. Such an
excess would be definitely higher than in the diffuse ISM, and should be
possibly higher than that in regions of ongoing star formation (age
$\lesssim 5 - 10$\,Myr) because the enrichment of large grains has not
started yet or has not lasted long enough to produce noticeable
effects.

UV observations of early-type stars in this region reaching down to
wavelengths of $\sim 1\,200 - 1\,500$\,\AA\, are needed to probe the
distribution of small  grains and understand whether the additional
component that we have detected is only composed of large grains or
whether smaller grains are also present and in which proportion. These
observations are possible with the {\em Cosmic Origin Spectrograph} on
board the HST. 

\section{Reddening distribution}

Since we have derived the extinction law characteristic of this  region,
we can measure the extinction towards any object in the field  whose
nominal location in the CMD is known. This is the case of the RC stars
and of those in the  upper MS (UMS). 

Zaritsky (1999) used observations from the Magellanic Clouds Photometric
Survey (Harris, Zaritsky \& Thompson 1997) to study the reddening
distribution towards hot ($T_{\rm eff} > 12\,000$\,K) and cold
(5\,500\,K $< T_{\rm eff} <$ 6\,500\,K) stars, finding that the former
appear on average more extinguished than the latter. However, in
Paper\,I we showed that this is not always the case: in the field that
we studied, located about $6\arcmin$ SW of 30\,Dor, UMS stars span a
narrower range of $E(B-V)$ values than RC objects. A similar conclusion
has been reached by Sabbi et al. (2013) in their preliminary study of
the Tarantula nebula. In the following we explore the distribution of
the absorbing material in the core of 30\,Dor, using the information
contained in the optical colour--colour (CC) and colour--magnitude
diagrams. We do not consider NIR bands here because our NIR  photometry
covers a substantially smaller field and because the lower NIR
extinction can be measured in fewer stars.

\begin{figure}
\centering
\resizebox{\hsize}{!}{\includegraphics[width=16cm]{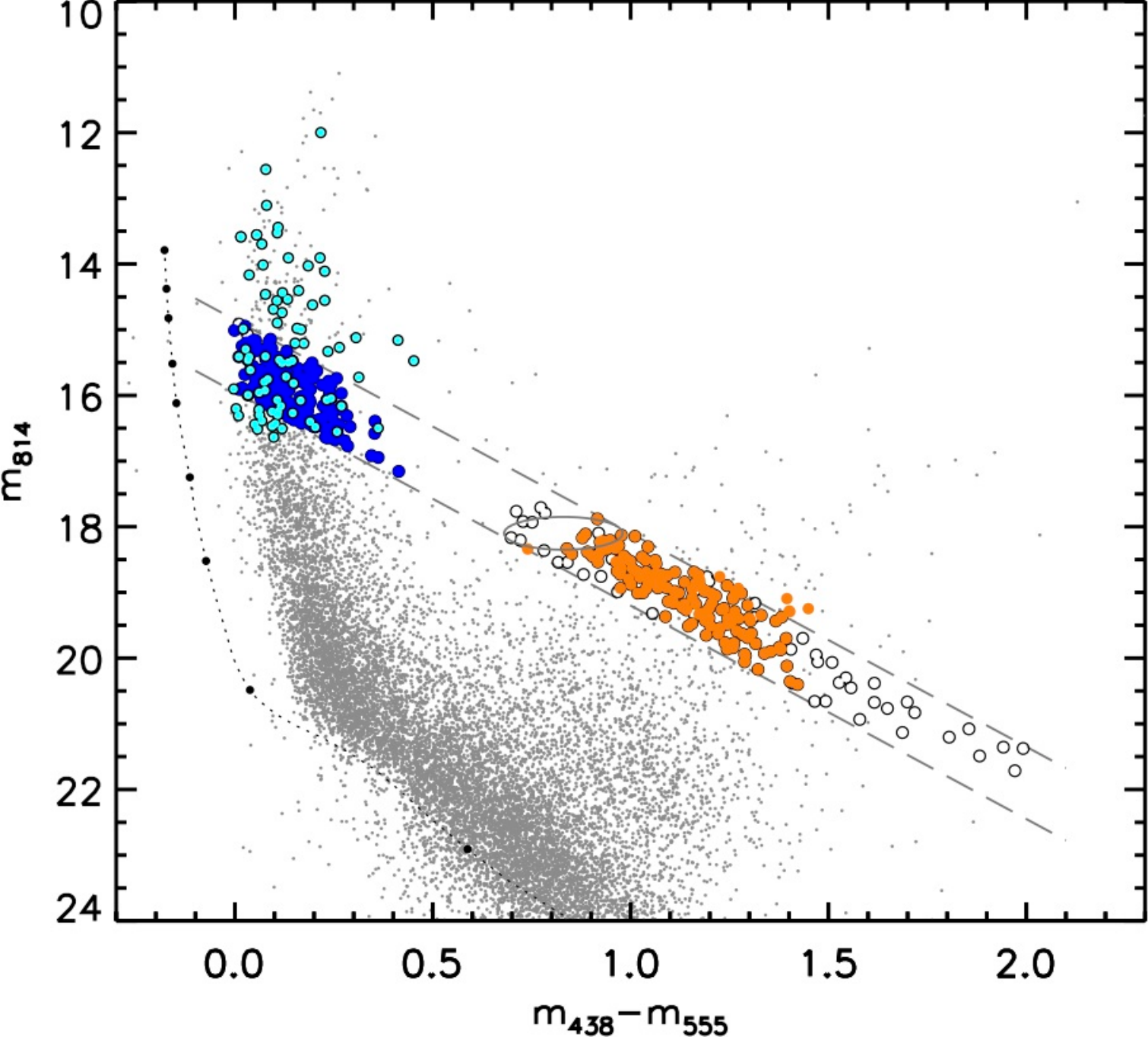}}
\caption{CMD showing selected UMS and RC stars for the study of the 
properties of the absorbing material.  All objects indicated as thick
dots have $\delta_3 < 0.1$. Those marked in blue and  orange have even
more accurate photometry in all optical bands ($\delta_4 < 0.1$). The
stars indicated in cyan are those in the sample of Ma\'{\i}z Apell\'aniz
et al. (2014). The  dashed lines are tangent to the spread ellipse
around the nominal RC location (also shown) and parallel to the
reddening vector. The ZAMS of Marigo et al. (2008) for $Z=0.007$ is
shown by the dotted line and already includes the foreground Galactic
extinction.  The points corresponding to masses of 1, 2, 5, 9, 15, 20,
30, 40, and 60\,\Msolar\, are indicated. }
\label{fig6}
\end{figure}

\begin{figure}
\centering
\resizebox{\hsize}{!}{\includegraphics{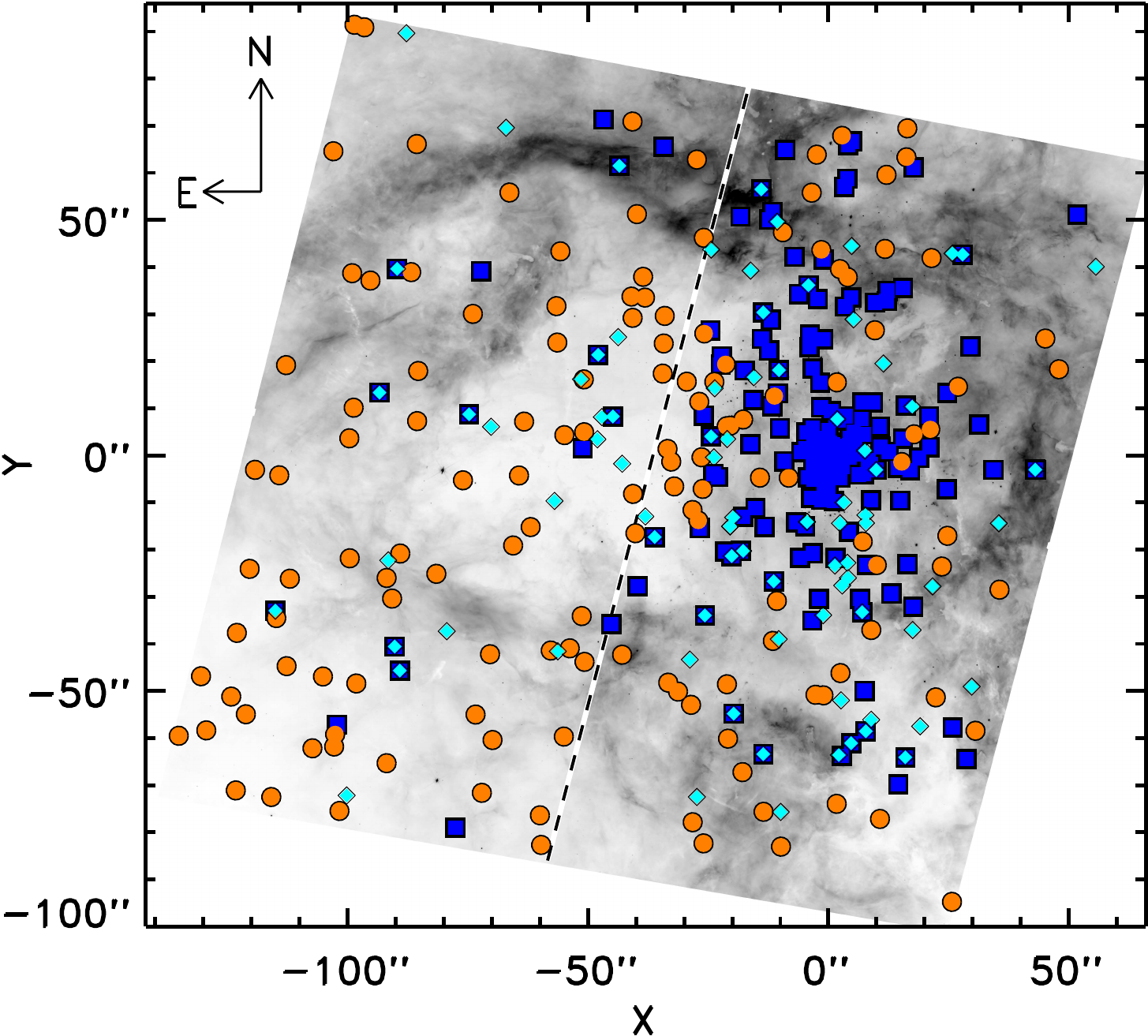}}
\caption{Location of the selected UMS and RC stars across the field. 
The objects in orange and blue are, respectively, the RC and UMS stars
shown in Figure\,\ref{fig6}. The objects in cyan are the 83 O-type and
B-type stars studied spectroscopically by Ma{\'i}z~Apell{\'a}niz et al.
(2014). The dashed line divides the region in two parts, the eastern and
western half-fields, each containing half (70) of the bona-fide RC
stars.}
\label{fig7}
\end{figure}

The objects that we have used for this study are indicated as thick
symbols in the $m_{814}$ {\em vs.} $m_{438}-m_{555}$ CMD of
Figure\,\ref{fig6}. They include the 140 bona-fide RC stars (shown in
orange) and an additional 202 UMS stars (blue) selected along the
extension of the reddening vector defined by the RC. The dashed lines in
the figure are tangent to the spread ellipse  around the nominal RC
location (also shown) and parallel to the reddening vector. Like in the
case of the RC stars, the UMS objects were selected to have a combined
photometric uncertainty in all optical bands of $0.1$\,mag or less
(i.e., $\delta_4 < 0.1$) and $W_{\rm eq}(H\alpha) < 3$\,\AA. The objects
marked in cyan in the figure are the 83 stars studied spectroscopically
by Ma{\'i}z~Apell{\'a}niz et al. (2014) that we will later compare to
our UMS stars. All blue objects in our  and Ma{\'i}z~Apell{\'a}niz et
al.'s sample have measured colours $B-V \ge 0$. The spatial location  of
all these objects is shown in Figure\,\ref{fig7}. 

In Section\,\ref{sec_ccd} we will discuss the reddening distribution
towards this sample of UMS and RC stars using the CC diagram. 
Furthermore, in Section\,\ref{sec_cmd} we will add to this sample the
objects marked as thick circles in Figure\,\ref{fig6} (45 stars in
total). These RC stars are typically fainter than the rest (most have $I
\gtrsim 20$), and their photometric uncertainty in the $U$ band is
somewhat larger, typically $\sim 0.2$\,mag. Therefore, these objects do
not meet the condition $\delta_4 < 0.1$ that we have imposed to select
the bona-fide RC stars discussed in Section\,3. However, their combined
photometric uncertainty in the  $B$, $V$ and $I$ bands still fully meets
that condition, i.e. with $\delta_3 < 0.1$ these objects have excellent
photometry in these bands. Their inclusion in the sample will prove
crucial to study more extinguished RC stars that are otherwise not
easily detectable in our shallower $U$-band exposures.

\subsection{Reddening from the colour--colour plane}
\label{sec_ccd}

\begin{figure}
\centering
\resizebox{\hsize}{!}{\includegraphics{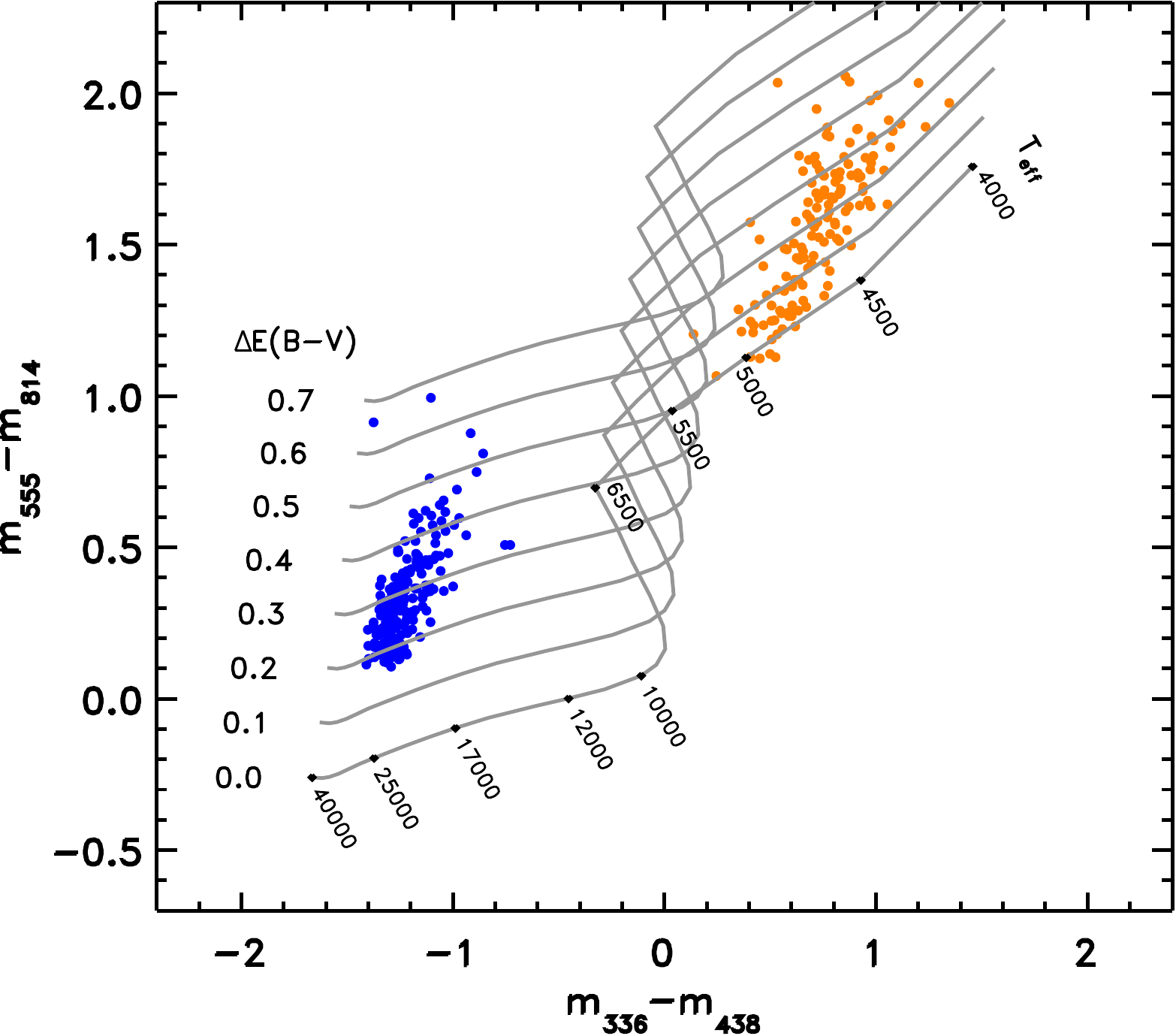}}
\caption{Colour--colour diagrams of the UMS and RC stars shown in blue
and orange, respectively, in Figure\,\ref{fig6} and \ref{fig7}. The
solid  curves provide the theoretical colours from the models of Bessell
et al. (1998) for stars with gravity $\log g=4.5$, metallicity $Z=0.007$
and effective temperature in the range 4\,000\,K -- 40\,000\,K, as
indicated. All curves already include the foreground MW extinction 
towards the LMC and each of them corresponds to a different level of
attenuation, indicated by the $\Delta E(B-V)$ value. }
\label{fig8}
\end{figure}

We begin by studying the reddening distribution in the CC diagram of
Figure\,\ref{fig8}, obtained by combining all available optical bands.
For now, we consider only stars  with $\delta_4 < 0.1$, namely the 140
bona-fide RC stars and the 202 UMS objects shown in Figure\,\ref{fig6},
respectively, in blue and orange  (the colours of the symbols stay the
same). The ``S''-shaped solid curve at the bottom of Figure\,\ref{fig8}
(marked $0.0$) corresponds to the theoretical colours from the models of
Bessell, Castelli \& Plez (1998) for stars with gravity $\log g=4.5$,
metallicity $Z=0.007$ and effective temperature in the range 4\,000\,K
-- 40\,000\,K, for the specific WFC\,3 bands used here (the values of
the  effective temperature are indicated along the curve). The curve
already includes the foreground MW contribution to the reddening, i.e.
$E(B-V)=0.07$.

The S-curve moves in the CC plane according to the attenuation caused by
the extinction in each band. As we discussed in Paper\,I, although
$R_\lambda$ is fixed by the extinction law at a specific wavelength, 
for a given $E(B-V)$ the observed extinction value $A_\lambda$ depends
on the spectrum of the star, because it is integrated over the entire
filter bandwidth. This effect is more pronounced for cold stars than for
hot stars and for blue filters than for red filters (see Romaniello et
al. 2002; Girardi et al. 2008). To produce the other S-curves of
Figure\,\ref{fig8}, with $\Delta E(B-V)>0$, we have applied to the
models of Bessell et al. (1998) mentioned above an attenuation
corresponding to incremental values of $E(B-V)$, assuming the extinction
specific to this region as derived in the previous section. We have
calculated the attenuation over the entire wavelength range covered by
the models, for values of $\Delta E(B-V)$ ranging from 0 to 1 in steps
of $0.01$ mag. Using {\em Synphot} (Laidler et al. 2005), we have folded
the attenuated models through the specific WFC\,3 bands and derived the
corresponding magnitudes in the HST system. The various S-curves in
Figure\,\ref{fig8} correspond to values of $\Delta E(B-V)$ ranging from
0 to $0.7$ in steps of $0.1$, as indicated.

\begin{figure}
\centering
\resizebox{\hsize}{!}{\includegraphics{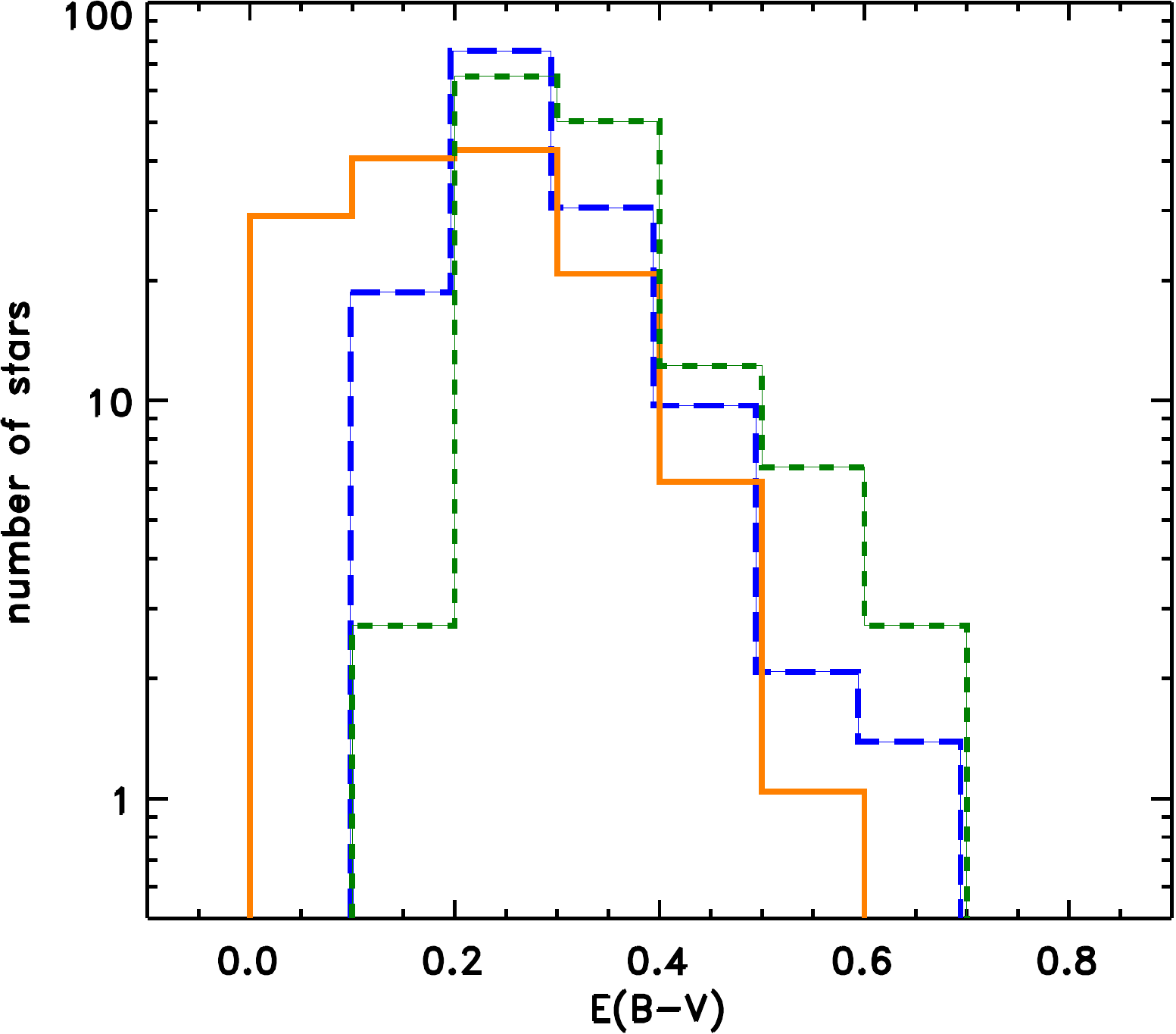}}
\caption{Histograms of the $E(B-V)$ distributions towards UMS stars
(blue dashed lines) and RC stars (orange solid lines). The three panels
correspond to those of Figure\,\ref{fig8}. The green  short-dashed line
in panel a) is the histogram of the $E(B-V)$ distribution measured by
Ma{\'i}z~Apell{\'a}niz et al. (2014) in this field. }
\label{fig9}
\end{figure}

Inspection of Figure\,\ref{fig8} reveals that UMS stars span a narrower
range of $\Delta E(B-V)$ values than RC objects, and all of them have
$\Delta E(B-V)> 0.18$. This difference is also seen in
Figure\,\ref{fig9}, showing separately the histograms of the
distribution of RC stars (orange solid line) and UMS stars (blue
long-dashed line), with the latter rescaled to the same number of stars
as RC objects. A smaller spread of $E(B-V)$ for UMS stars is not
surprising, because the UMS stars are associated with R\,136 and have a
limited spatial distribution along the line of sight.

The $\Delta E(B-V)$ values that we have derived for the UMS stars can be
compared with with those derived spectroscopically in the same field by
Ma{\'i}z~Apell{\'a}niz et  al. (2014). These authors have analysed a
total of 67 O-type stars and 16 B-type stars, for which spectra are
available in the range $\sim 4\,000 - 5\,000$\,\AA\ at a spectral
resolution $R \simeq 8\,000$ from the ``VLT--FLAMES Tarantula  Survey''
(Evans et al. 2011). These stars are indicated as cyan dots in
Figures\,\ref{fig6} and \ref{fig7}. Having determined the spectral types
for these objects allows Ma{\'i}z~Apell{\'a}niz et al. (2014) to derive
an accurate value of the extinction towards each of them. The green 
short-dashed line in Figure\,\ref{fig9} shows the histogram of their
$E(B-V)$ distribution. For ease of comparison with our results, we have
shifted the histogram of Ma{\'i}z~Apell{\'a}niz et al. (2014) by
$0.07$\,mag to the left, to account for the foreground MW extinction,
since that is not included in our $\Delta E(B-V)$ values. The reddening
distribution is very similar to the one we found. The only statistically
significant difference is in the number of stars with $\Delta E(B-V) =
0.15$ and we attribute this to the different composition of the two
samples, since ours is exclusively made up of late O-type stars in the
range $20 - 40$\,\Msolar, while the objects in Ma{\'i}z~Apell{\'a}niz et
al.'s (2014) study range from early O-type stars to early B-type objects
(see Figure\,\ref{fig6}). 

From this comparison we conclude that the $E(B-V)$ values that we derive
are reliable, also for UMS stars, confirming that for young massive
stars our method is a solid and relatively simple way to study the
amount and properties of the extinction from photometry alone, with no
need for spectroscopy. Provided that there are enough RC stars in the
field and the level of extinction is high and variable, our method
allows us to derive a robust extinction law and the absolute value of
the extinction towards most UMS and RC stars.

\subsection{Reddening from the colour--magnitude plane}
\label{sec_cmd}

The histograms in Figure\,\ref{fig9} also appear to indicate that,
within the uncertainties, the maximum extinction value is substantially
the same for UMS and RC stars. However, we already know that some of the
most extinguished RC objects are missing in our $U$-band photometry
because it is shallower than in the other bands, thereby limiting the
spread in the measured distribution. As discussed in Section\,3, after
an initial selection of candidates in each CMD, we retained as bona-fide
RC stars only those classified as such in  all CMDs simultaneously. This
is a necessary step to ensure that the derived extinction law is robust
and accurate at all wavelengths, but it implies that the band with
shallowest exposures ($U$ in our case) limits the most extinguished and
hence faintest RC stars that we can retain in our sample. On the other
hand, since we have already derived the extinction law, for the study of
the reddening distribution we can rely on the CMDs based on the deepest
exposures. 

To this end, we have used the $I$ {\em vs.} $B-V$ CMD of
Figure\,\ref{fig6}, since in these bands the exposures are deeper and
still sample a relatively wide spectral range ($\sim 4\,000 -
9\,000$\,\AA). We have selected all the objects inside the strip shown
in Figure\,\ref{fig6}, parallel to the reddening vector in this plane,
with combined photometric uncertainty in the three bands $\delta_3 <
0.1$. As before, we have excluded from the sample all stars with
$W_{\rm eq}(H\alpha) > 3$\,\AA. There are a total of 381 objects
satisfying these conditions: the 202 UMS stars discussed above and 179
objects that have $B-V > 0.7$ and are likely to be RC stars. Most of
them are the bona-fide RC objects discussed above, to which we have
added the 45 stars indicated as thick circles in Figure\,\ref{fig6}.

\begin{figure}
\centering
\resizebox{\hsize}{!}{\includegraphics{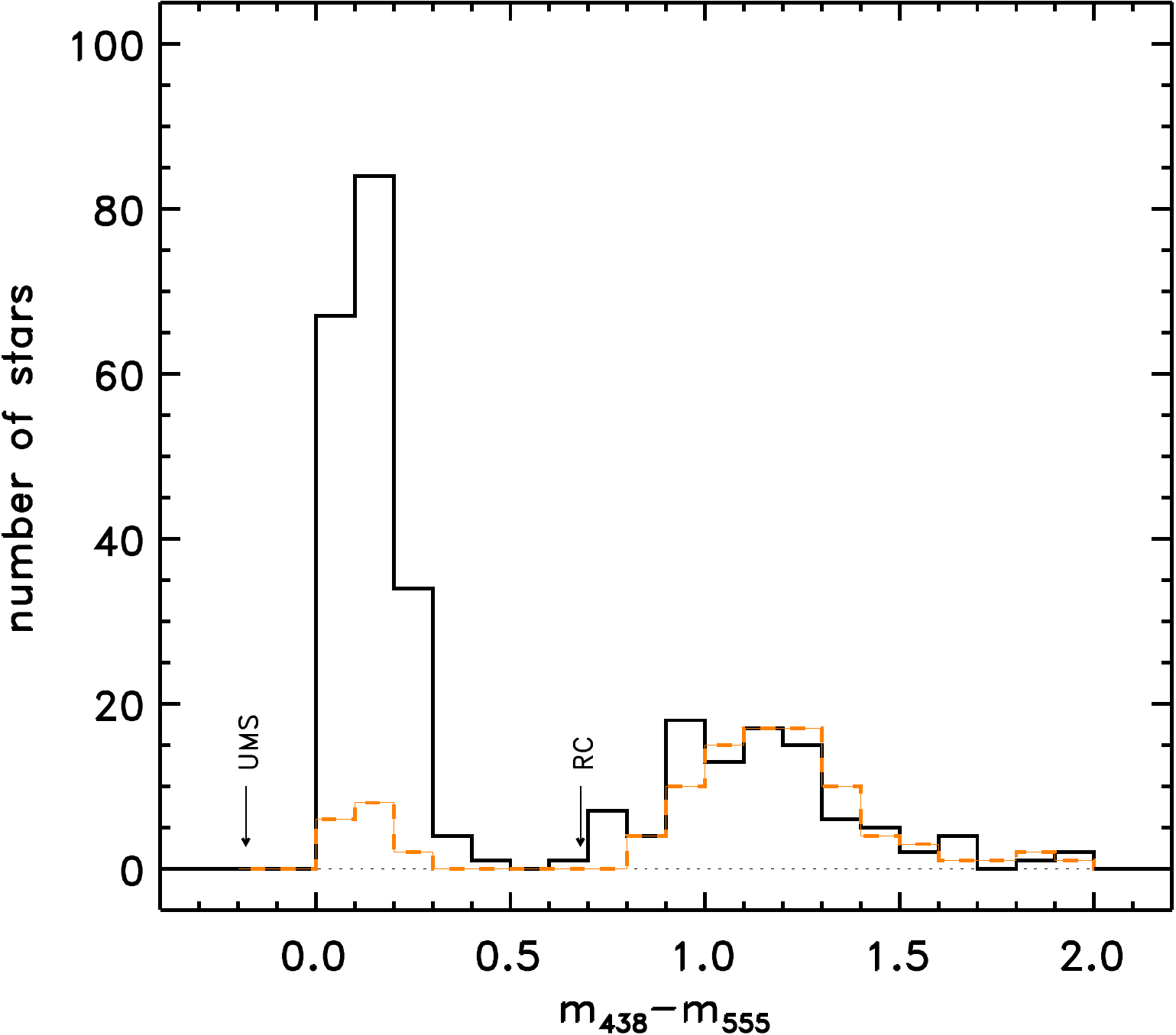}}
\caption{Histograms of the observed $m_{438}-m_{555}$ colours of the UMS
and RC stars contained inside the dashed lines in Figure\,\ref{fig6}.
The solid line corresponds to the stars in the western half-field and
the dashed line to those in the eastern half-field. The arrows
indicate the bluest expected colours of an UMS or RC stars, taking into
account the $E(B-V)=0.07$ intervening Galactic extinction.}
\label{fig10}
\end{figure}

The histograms shown in Figure\,\ref{fig10} provide the distribution of
the observed  $m_{438}-m_{555}$  colours of these 381
stars.\footnote{Although the $m_{438}$ and $m_{555}$ bands do not
exactly coincide with the Johnson $B$ and $V$ bands, the difference is
not significant in the context of the discussion that follows and we
will ignore it.} The solid line corresponds to the stars in the western
half-field, where R\,136 is located, while the dashed line is for the
objects in the eastern half-field. The arrows indicate the
bluest expected colours of an UMS or RC stars after the $E(B-V)=0.07$
intervening MW extinction is taken into account. 

Figure\,\ref{fig10} confirms that in all cases the $\Delta E(B-V)$
spread, calculated from the vertical arrows, is larger for RC stars than
for UMS stars, as we already concluded from Figure\,\ref{fig9}. In
addition, here there are systematically more highly extinguished RC
stars, which were not detected in the shallower $U$ band exposures and
thus did not appear in Figure\,\ref{fig9}. Zaritsky (1999) had concluded
that, on average, in the LMC young stars are more obscured by dust than
old stars, but this conclusion too was based on a sample of stars
detected simultaneously in the $U$, $B$, $V$, and $I$ bands. As
mentioned above and as already pointed out by Sabbi et al. (2013), this
selection criterion penalizes the most extinguished evolved stars,
thereby resulting in an incomplete reddening statistics. 

The presence of more extinguished RC stars than UMS objects has
important implications for the distribution of the absorbing material. 
In particular, the UMS stars in our field are associated with R\,136:
they have a limited spatial distribution along the line of sight and
their main source of extinction is the molecular cloud from which R\,136
formed. The presence of more extinguished RC stars indicates that the
absorbing material is present along the line of sight also well behind
R\,136. Also in the eastern half-field (orange dashed line) are the RC
stars more heavily reddened than UMS objects, but an important
difference with respect to the western half-field is that here the
minimum $\Delta E(B-V)$ value towards RC stars is $\sim 0.2$ larger.
This implies that there is an additional extinction component in front
of the eastern half-field, roughly coinciding with the silhouette of a
``Christmas tree'' in Figure\,\ref{fig1}.

The most straightforward explanation is that the absorbing material in
the Christmas tree area is an outflow associated with a previous star
formation episode in that area, some $\sim 50$\,Myr ago. The outflow has
been driven outward, towards the observer, by winds and supernova
explosions from former massive stars through an opening in the confining
molecular cloud. Indeed, the region of the Christmas tree coincides
with a blue-shifted outflow with a velocity of 100\,km\,s$^{-1}$ (Chu \&
Kennicutt 1994), as one would expect from stellar winds and supernova
explosions.The average density of the material and the associated column
density are lower than in the original star forming region, as confirmed
by the low emission at IR (e.g. Indebetouw et al. 2009) and radio
wavelengths (e.g. Johansson et al. 1998; Rubio et al. 1998). The outflow
extends in our direction and causes an appreciable extinction in the
foreground. 

Velocities in excess of 10\,km\,s$^{-1}$ are typical of the outflows
powered by massive stars in Orion ({\em e.g.,} Bally 2007), thus in
50\,Myr or so this absorbing component has had time to extend few
hundred parsec towards the observer and now affects also RC stars in the
foreground of 30\,Dor. Conversely, RC stars in the foreground of R\,136,
in the western half-field, are not affected by this material and the
extinction towards them is lower. Previous HST observations (De Marchi
et al. 2011b) have revealed a conspicuous population of $\sim 30$\,Myr
old PMS stars in the eastern half-field, indicating that star formation
was indeed ongoing in this area well before the birth of the R\,136
cluster. It is the $> 8$\,\Msolar\, stars belonging to that generation
that were responsible for the injection of larger grains in the ISM when
they exploded as supernovae.

\begin{figure}
\centering
\resizebox{\hsize}{!}{\includegraphics{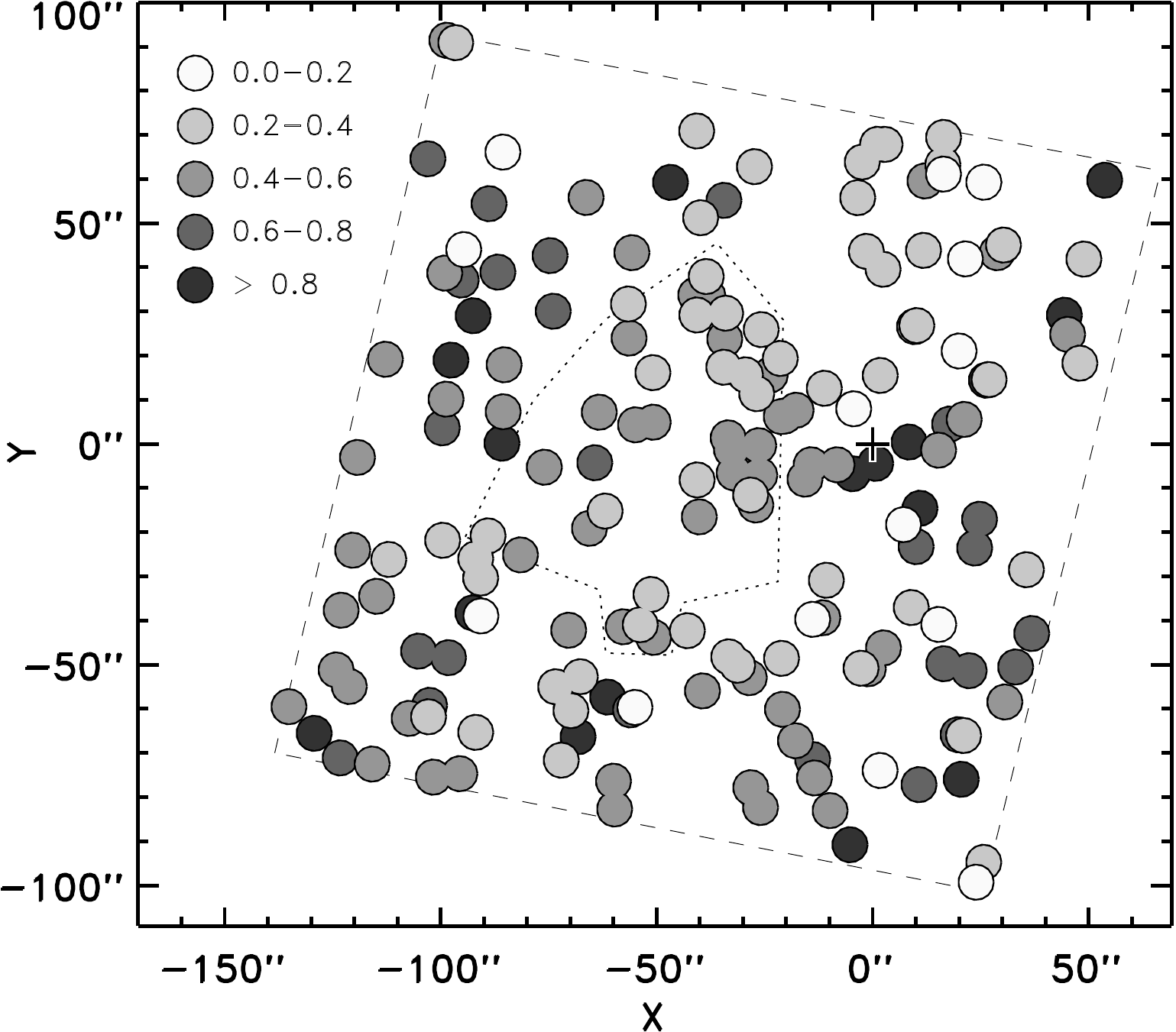}}
\caption{Distribution of the RC stars across the entire field.
Progressively darker shades of grey correspond to progressively higher
$E(B-V)$ values, as per the legend. The area of the Christmas tree
contains no stars with $E(B-V) < 0.2$.}
\label{fig11}
\end{figure}

The absence of objects with $E(B-V) < 0.2$ inside the Christmas-tree region
is evident in Figure\,\ref{fig11}, showing the spatial distribution of
the RC stars across the entire field with progressively darker shades of
grey. Although there are 37 RC stars inside the area outlined by the
dotted line, or 21\,\% of the total, none of them have $E(B-V) < 0.2$.
In this region also stars with $E(B-V) > 0.6$ are missing, indicating 
that the total column density is lower than elsewhere in the field, in
excellent agreement with the evacuated region revealed by the X ray
observations of Townsley et al. (2006) in this area.  In general,
Figure\,\ref{fig11} provides a map of the column density towards all RC
stars and, if the spatial density of these objects is known or is
assumed to be uniform, it can be used to constrain the distribution of
the absorbing material along the line of sight. This will be the topic
of a forthcoming paper (Panagia \& De Marchi, in preparation).

\section{Summary and conclusions}

Using WFC\,3 observations of the core of 30\,Dor (De Marchi et al.
2011a), we have studied the properties of interstellar extinction  in
the $U, \, B, \, V, \, I, \, J,$ and $H$ bands over a field of $2\farcm7
\times 2\farcm7$ including the central R\,136 cluster. We have taken
advantage of the considerable and uneven levels of extinction in this
field, which  affect the colours and magnitudes of RC stars and spread
them across the CMD. Following the method developed in Paper\,I, we have
derived in an accurate and quantitative way both the extinction law 
$R_\lambda$ in the wavelength range $\sim 0.3 - 1.6\,\muup$m and the
values of the absolute extinction $A_\lambda / A_V$ towards $\sim 180$
RC objects as well as $\sim 200$ O-type stars. The main results of this
work can be summarised as follows.

\begin{enumerate}

\item  

Comparing our observations in all bands with the theoretical colours and
magnitudes of RC stars, we define the region of the CMDs where reddening
can place RC stars. In this region we find a total of 146 objects
detected in all optical bands and with a combined photometric
uncertainty $\delta_4 < 0.1$\,mag, 93 of which are also detected in the
NIR bands covering a smaller area. We exclude from this sample 6 stars
(of which 3 in the NIR) that have $W_{\rm eq}(H\alpha) > 3$\,\AA\, as
they could be contaminating PMS stars. The resulting sample of bona-fide
RC stars includes 140 objects, of which 90 also covered by the NIR
observations.

\item 

We derive the best linear fit to the distribution of the bona-fide RC
stars in all CMDs, obtaining in this way the absolute extinction and 
the ratio $R$ between absolute and selective extinction in the specific 
WFC\,3 bands. Interpolation at the wavelengths of the Johnson $B$ and
$V$ bands provides the extinction curve in the canonical form
$R_\lambda  \equiv A_\lambda / E(B-V)$, in the range $\sim 0.3 -
1.6\,\muup$m. 

\item

With $R_V=4.5 \pm 0.2$, the extinction law that we measure in R\,136 is
considerably flatter than that of the diffuse Galactic ISM ($R_V=3.1$).
Furthermore, while the laws by Cardelli et al. (1989), Fitzpatrick \&
Massa (1990), and Ma\'{\i}z Apell\'aniz et al. (2014) may provide a good
fit to our extinction curve at optical wavelengths with $R_V$ suitably
close to $4.5$, none of them offers a good fit to the NIR observations.
Instead, the extinction curve of R\,136 is best represented at optical
wavelengths by the Galactic curve shifted vertically by an offset of
$1.5$. Longwards of $1\,\muup$m the extinction is very well matched by
the Galactic law multiplied by a factor of $2.2$. This implies that in 
R\,136 the relative fraction of large grains is a factor of $2.2$ higher.

\item

We interpret the observed extinction curve as the result of an excess
component of large grains added to the canonical grain distribution
typical of the diffuse Galactic ISM. This is consistent with the 
selective injection by Type II supernova explosions of ``fresh'' large
grains into a MW mix, as recently revealed by {\em Herschel} and {\em
ALMA} observations of SN\,1987A.

\item   

Having derived the extinction curve in R\,136, we study the relative
distribution of stars and dust in the field by comparing the extinction
towards individual RC and UMS stars.  The $E(B-V)$ values measured
towards 179 uniformly distributed RC stars provide a map of the column
density in the region. The map is richly populated, since in the typical
Magellanic  Clouds HST field we can count on a surface density of 20
such stars per arcmin$^2$. We also study the extinction towards 202 UMS
stars and find that they span a narrower range of $E(B-V)$ values than
RC stars. The same result was found in Paper\,I in a field located
$6\arcmin$ SW of the cluster. This is at odds with the conclusions of 
Zaritsky (1999), who found that in the LMC stars with $T_{\rm eff} <
12\,000$\,K are typically affected by larger extinction than stars with
$T_{\rm eff} = 5\,500 - 6\,500$\,K. We conclude that the apparent
absence of heavily reddened evolved stars in Zaritsky's (1999) sample
was caused by a selection effect, namely by his exclusion of hard to
detect stars from the $U$-band photometry. 

\item 

We compare the $E(B-V)$ values obtained with our method with those
derived spectroscopically by Ma\'{\i}z Apell\'aniz et al. (2014) for a
sample of 83 early type stars in the field. We find an excellent
agreement, confirming the validity of our extinction determination not
only for RC stars but also for O-type and B-type stars that are still on
the MS.

\end{enumerate}

In summary, in this work we have shown that when the levels of
extinction are high and uneven, like in the case of the 30\,Dor region,
RC stars can be used to derive the extinction law and the absolute
extinction towards those stars with an accuracy comparable with that
allowed by spectroscopy of early-type stars. Furthermore, since RC stars
sample in a uniform way a much larger volume than early-type stars, they
can provide a characterisation of the extinction and a measure of the
column density also outside the most active star forming regions and
over multiple lines of sight.

We have also shown that in the range $\sim 0.3 - 1.6\,\muup$m the
extinction curve of R\,136 is not satisfactorily described by the widely
used single-parameter families of extinction laws (e.g. Cardelli et al.
1989; Fitzpatrick \& Massa 1990). {This is not unexpected, since
Gordon et al. (2003) had already shown that the the extinction curve in
the LMC2 Supershell around 30\,Dor and along most lines of sight in the
LMC do not follow the Cardelli et al. (1989) relationship based on
$R_V$.} In principle, it is practical to have a parametric description,
but the extinction law is governed by the interplay of scattering and
absorption due to grains with different physical properties, including
chemical composition and size, and located in different environments.
Therefore, one wonders whether it is possible and meaningful to look for
a mathematical formulation able to capture the complexity revealed by
the observations, particularly when one needs to cover a wide spectral
range. 

Instead, it is very important to study the properties of the extinction
curve in light of the current and recent star formation history
experienced by the regions. So far our investigation using this method
has covered two fields in the LMC, one containing R\,136 and one located
$6\arcmin$ SW of it. We are going to extend the study to the entire
Tarantula nebula, which was recently observed as part of the Hubble
Tarantula Treasury Program (Sabbi et al. 2013). These observations
sample a large contiguous area on the sky (168 arcmin$^2$) in the range 
$\sim 0.25 - 1.6\,\muup$m and will allow us to also explore how smaller
grains contribute to the extinction in regions with different intensity
of the current and most recent star formation.

\section*{Acknowledgments}

We are grateful to our referee, Prof. Geoff Clayton, for insightful
comments that have helped us to improve the presentation of this work.
This paper is based on Early Release Science observations made by the
WFC\,3 Scientific Oversight Committee. NP acknowledges partial support
by STScI--DDRF grant D0001.82435.


\begin{thebibliography}{References}

\bibitem[]{} Bally J., 2007, Ap\&SS, 311, 15 

\bibitem[]{} Bessell M., Castelli F., Plez B., 1998, A\&A, 333, 231

\bibitem[]{} Bless R., Savage B., 1972, ApJ, 171, 293

\bibitem[]{} Cardelli J., Clayton G., 1988, AJ, 95, 516

\bibitem[]{} Cardelli J., Clayton G., Mathis J., 1988, ApJ, 329, L33

\bibitem[]{} Cardelli J., Clayton G., Mathis J., 1989, ApJ, 345, 245

\bibitem[]{} Cardelli J., Sembach K., Mathis J., 1992, AJ, 104, 1916

\bibitem[]{} Chu Y.-H., Kennicutt R., 1994, ApJ, 425, 720

\bibitem[]{} Clayton G., Martin P., 1985, ApJ, 288, 558

\bibitem[]{} De Marchi G., et al., 2011a, ApJ, 739, 27

\bibitem[]{} De Marchi G., et al., 2011b, ApJ, 740, 11

\bibitem[]{} De Marchi G., Panagia N., Girardi L., 2014, MNRAS, 438, 513

\bibitem[]{} De Marchi G., Panagia N., Romaniello M., 2010, ApJ, 715, 1

\bibitem[]{} De Marchi G., Panagia N., Sabbi E., 2011, ApJ, 740, 10 

\bibitem[]{} Draine B., Lee H., 1984, ApJ, 285, 89 

\bibitem[]{} Evans C., et al., 2011, A\&A, 530, A108

\bibitem[]{} Fitzpatrick E., 1998, in Ultraviolet Astrophysics Beyond 
  the IUE Final Archive, ed. W. Wamsteker, R. Gonzalez Riestra 
  (Noordwijk: ESA), 461

\bibitem[]{} Fitzpatrick E., 1999, PASP, 111, 63 

\bibitem[]{} Fitzpatrick E., Massa D., 1990, ApJS, 72, 163

\bibitem[]{} Fitzpatrick E., Massa D., 1999, ApJ, 525, 1011

\bibitem[]{} Fitzpatrick E., Massa D., 2005, AJ, 130, 1127 

\bibitem[]{} Fitzpatrick E., Massa D., 2007, ApJ, 663, 320

\bibitem[]{} Fitzpatrick E., Savage B., 1984, ApJ, 279, 578 

\bibitem[]{} Gall C., et al., 2014, Nature, 511, 326

\bibitem[]{} Geha M., et al., 1998, AJ, 115, 1045

\bibitem[]{} Girardi L., et al., 2008, PASP, 120, 583 

\bibitem[]{} Gordon K., Clayton G., Misselt K., Landolt A., 
  Wolff M., 2003, ApJ, 594, 279

\bibitem[]{} Gottlieb D., Upson W., 1969, ApJ, 157, 611 

\bibitem[]{} Greenberg J. M., 1968, in Nebulae and interstellar matter, ed. 
  B. Middlehurst, L. Aller (Chicago: Univ. Chicago Press), 221

\bibitem[]{} Harris J., Zaritsky D., Thompson I., 1997, AJ, 114, 1933 

\bibitem[]{} Haschke R., Grebel E., Duffau S., 2011, AJ, 141, 158 

\bibitem[]{} Heckman T., et al., 2004, ApJ, 613, 109

\bibitem[]{} Hill V., Andrievsky S., Spite M., 1995, A\&A, 293, 347

\bibitem[]{} Hunter D., et al., 1995, ApJ, 444, 758

\bibitem[]{} Indebetouw R., et al., 2009, ApJ, 694, 84

\bibitem[]{} Indebetouw R., et al., 2014, ApJ, 782, L2 

\bibitem[]{} Johansson L., et al. 1998, A\&A, 331, 857

\bibitem[]{} Johnson H., 1967, ApJ, 150, L39

\bibitem[]{} Johnson H., Mendoza E., 1964, Boletin de los Obs. de 
  Tonantzintla y Tacubaya, 3, 311

\bibitem[]{} Johnson H., 1968, in  Nebulae and interstellar matter, ed. 
  B. Middlehurst, L. Aller (Chicago: Univ. Chicago Press), 167

\bibitem[]{} Laidler V., et al, 2005, Synphot User's Guide, (Baltimore:
  STScI)

\bibitem[]{} Lebouteiller V., Bernard-Salas J., Brandl B., et al.,
   2008, ApJ, 680, 398 

\bibitem[]{} Lee T., 1968, ApJ, 152, 913

\bibitem[]{} Lilly S., Le Fevre O., Hammer F., Crampton D., 1996, ApJ,
  460, L1

\bibitem[]{} Madau P., et al., 1996, MNRAS, 283, 1388

\bibitem[]{} Ma\'{\i}z Apell\'aniz J., 2004, PASP, 116, 859

\bibitem[]{} Ma\'{\i}z Apell\'aniz J., et al., 2014, A\&A, 564, A63 

\bibitem[]{} Marigo P., et al., 2008, A\&A, 482, 883 

\bibitem[]{} Massa D., Savage B., Fitzpatrick E., 1983, ApJ, 266, 662

\bibitem[]{} Mathis J., Rumpl W., Nordsieck K., 1977, ApJ, 217, 425 

\bibitem[]{} Matsuura M., et al., 2011, Sci, 333, 1258

\bibitem[]{} McCall M., 2004, AJ, 128, 2144

\bibitem[]{} Meurer G., Heckman T., Lehnert M., Leitherer C., 
  Lowenthal J., 1997, AJ, 114, 54 

\bibitem[]{} Misselt K., Clayton G., Gordon K., 1999, ApJ, 512, 128

\bibitem[]{} Panagia N., 1999, in New Views of the Magellanic Clouds,
  ed. Y.-H. Chu, N. Suntzeff, J. Hesser, D. Bohlender (San Francisco: 
  ASP), 549 

\bibitem[]{} Panagia N., Gilmozzi R., Macchetto F., Adorf H.-M., 
  Kirshner R., 1991, ApJ, 380, L23 

\bibitem[]{} Romaniello M., 1998, PhD thesis, Scuola Normale Superiore,
Pisa, Italy

\bibitem[]{} Romaniello M., Panagia N., Scuderi S., Kirshner R., 
  2002, AJ, 123, 915

\bibitem[]{} Rubio M., et al., 1998, AJ, 116, 1708

\bibitem[]{} Sabbi E., et al., 2013, AJ, 146, 53

\bibitem[]{} Seaton M., 1979, MNRAS, 187, 73P

\bibitem[]{} Shapley A., Steidel C., Pettini M., Adelberger K., 
  2003, ApJ, 588, 65 

\bibitem[]{} Savage B., Mathis J., 1979, ARA\&A, 17, 73 

\bibitem[]{} Tatton B., et al., 2013, A\&A, 554, A33 

\bibitem[]{} Townsley L., et al., 2006, AJ, 131, 2140

\bibitem[]{} Valencic L., Clayton G., Gordon K., 2004, ApJ, 616, 912 

\bibitem[]{} van de Hulst H., 1957, Light scattering by small particles,
  (New York: Wiley \& Sons)

\bibitem[]{} Walborn N., Blades J., 1997, ApJS, 112, 457 

\bibitem[]{} White R., Basri G., 2003, ApJ, 582, 1109 

\bibitem[]{} Zaritsky D., 1999, AJ, 118, 2824 


\end{thebibliography}
\end{document}